\documentclass[12pt]{article} 

\usepackage{graphicx}
\usepackage{amsmath}
\usepackage{amssymb}
\usepackage{hyperref}
\usepackage[height=8.85in,width=6.45in]{geometry}
\usepackage{tikz}

\numberwithin{equation}{section}

\newcommand{\eg}{\textit{e.g.}}
\newcommand{\ie}{\textit{i.e.}}
\newcommand{\etc}{\textit{etc.}}

\newcommand{\be}{\begin{equation}}
\newcommand{\ee}{\end{equation}}
\newcommand{\bea}{\begin{equation}\begin{aligned}}
\newcommand{\eea}{\end{aligned}\end{equation}}
\newcommand{\beq}{\begin{equation}}
\newcommand{\eeq}{\end{equation}}
\newcommand{\beqa}{\begin{eqnarray}}
\newcommand{\eeqa}{\end{eqnarray}}

\newcommand{\smat}[1]{\big( \begin{smallmatrix} #1 \end{smallmatrix} \big)}
\newcommand{\parfrac}[2]{\frac{\partial #1}{\partial #2}}
\newcommand{\intd}[2]{{\raisebox{#1}{\scriptsize \ensuremath{#2}}}}

\newcommand{\cA}{\mathcal{A}}

\newcommand{\cC}{\mathcal{C}}

\newcommand{\cE}{\mathcal{E}}
\newcommand{\cF}{\mathcal{F}}

\newcommand{\cI}{\mathcal{I}}

\newcommand{\cL}{\mathcal{L}}
\newcommand{\cM}{\mathcal{M}}
\newcommand{\cN}{\mathcal{N}}
\newcommand{\cO}{\mathcal{O}}

\newcommand{\cQ}{\mathcal{Q}}

\newcommand{\bC}{\mathbb{C}}

\newcommand{\bI}{\mathbb{I}}

\newcommand{\bP}{\mathbb{P}}

\newcommand{\bR}{\mathbb{R}}
\newcommand{\bW}{\mathbb{W}}
\newcommand{\bZ}{\mathbb{Z}}

\newcommand{\fB}{\mathfrak{B}}

\newcommand{\fC}{\mathfrak{C}}

\newcommand{\fg}{\mathfrak{g}}
\newcommand{\fh}{\mathfrak{h}}
\newcommand{\fk}{\mathfrak{k}}

\newcommand{\fM}{\mathfrak{M}}
\newcommand{\fP}{\mathfrak{P}}

\newcommand{\fR}{\mathfrak{R}}

\newcommand{\fs}{\mathfrak{s}}

\newcommand{\sQ}{\mathsf{Q}}
\newcommand{\sT}{{\sf T}}

\newcommand{\e}{\mathrm{e}}
\newcommand{\g}{\mathrm{g}}
\newcommand{\dd}{\mathrm{d}}

\newcommand{\slj}{{/\hspace{-.40em}j}}
\newcommand{\sli}{{/\hspace{-.35em}i}}

\newcommand{\bi}{{\overline{\imath}}}
\newcommand{\bj}{{\overline{\jmath}}}

\newcommand{\R}{\bR}

\newcommand{\CC}{{\mathbb C}}
\newcommand{\PP}{{\mathbb P}}

\newcommand{\CP}{{\CC\PP}}

\newcommand{\ts}{\textstyle}

\newcommand{\rep}[1]{\ensuremath{\mathbf{#1}}}

\def\R{\bR}
\def\dbar{\bar\partial}
\newcommand{\Deps}{\Delta_{\varepsilon}}

\DeclareMathOperator{\Tr}{Tr}
\DeclareMathOperator{\Cone}{Cone}

\DeclareMathOperator{\sign}{sign}
\DeclareMathOperator{\rank}{rank}

\DeclareMathOperator{\im}{\bI m}
\DeclareMathOperator*{\Res}{Res}
\DeclareMathOperator*{\JKres}{JK-Res}

\newcommand{\nocontentsline}[3]{}
\let\addcontoriginal\addcontentsline
\newcommand{\tocless}[2]{\let\addcontentsline\nocontentsline #1{#2} \let\addcontentsline\addcontoriginal}

\begin{document}

\begin{titlepage}

\begin{flushright}
IPMU-13-0146\\
UT-13-29
\end{flushright}

\vskip 3cm

\begin{center}
{\Large \bf
Elliptic genera of 2d $\cN{=}2$ gauge theories
}

\vskip 2.0cm

Francesco Benini$^\sharp$,
Richard Eager$^\natural$,
Kentaro Hori$^\natural$,
and Yuji Tachikawa$^\flat$
\bigskip
\bigskip

\begin{tabular}{ll}
$^\sharp$&Simons Center for Geometry and Physics, Stony Brook University,\\
&Stony Brook, NY 11794, USA\\
$^\natural$  & Institute for the Physics and Mathematics of the Universe (WPI), \\
& University of Tokyo,  Kashiwa, Chiba 277-8583, Japan\\
$^\flat$  & Department of Physics, Faculty of Science, \\
& University of Tokyo,  Bunkyo-ku, Tokyo 133-0022, Japan
\end{tabular}

\vskip 1cm

\textbf{Abstract}
\end{center}

\medskip
\noindent
We compute the elliptic genera of general two-dimensional $\cN{=}(2,2)$ and $\cN{=}(0,2)$ gauge theories.
We find that the elliptic genus is given by the sum of Jeffrey-Kirwan residues of a meromorphic form, representing the one-loop determinant of fields, on the moduli space of flat connections on $T^2$.
We give several examples illustrating our formula, with both Abelian and non-Abelian gauge groups, and discuss some dualities for $U(k)$ and $SU(k)$ theories. This paper is a sequel to the authors' previous paper \cite{Benini:2013nda}.


\end{titlepage}

\setcounter{tocdepth}{2}
\tableofcontents


\section{Introduction}
\label{sec: intro}

Supersymmetric localization is a powerful technique that allows us to exactly compute the partition function of a supersymmetric theory on a supersymmetric background and the expectation values of certain operators. Our aim in this paper is to obtain a formula for the elliptic genera, \ie{}, the partition functions on $T^2$ with supersymmetric boundary conditions of $\cN{=}(2,2)$ and $\cN{=}(0,2)$ gauge theories in two dimensions.

Typically, the infinite-dimensional path integral of the field theory is reduced to an integral of the one-loop determinant over the finite-dimensional moduli space of supersymmetric (or BPS) configurations.
For our gauge theory on $T^2$ with gauge group $G$ of rank $r$, the moduli space of BPS configurations is the moduli space $\fM$ of flat connections of $G$ over $T^2$. It has real dimension $2r$.
Given a complex structure $\tau$ on $T^2$, $\fM$ inherits a natural complex structure making it an $r$-dimensional complex torus.  The one-loop factor $Z_\text{1-loop}$ is naturally a meromorphic $(r,0)$-form on $\fM$, and therefore it is natural to guess that the elliptic genus is given by a kind of residue operation formula
$$
Z_{T^2} = \frac1{(2\pi i)^r} \oint_{C} Z_\text{1-loop}
$$
where $C$ is an appropriate real $r$-dimensional cycle in $\fM$. The hard task is to find the correct cycle $C$.

In a previous paper \cite{Benini:2013nda}, the cycle $C$ has been determined when the gauge theory has rank one, \ie{}, $r=1$.
In that case, the poles  of $Z_\text{1-loop}$ can be split into two groups, $u\in \fM_\text{sing}^+$ and $u\in \fM_\text{sing}^-$, distinguished by the sign of the charges of the fields responsible for the divergence.
Then the formula is given by
\be
\label{rank1-schematic}
Z_{T^2} = \frac1{|W|} \sum_{u_+ \,\in\, \fM_\text{sing}^+} \frac1{2\pi i} \oint_{u=u_+} \hspace{-.2cm} Z_\text{1-loop}
= - \frac1{|W|} \sum_{u_- \,\in\, \fM_\text{sing}^-} \frac1{2\pi i} \oint_{u=u_-} \hspace{-.2cm} Z_\text{1-loop}
\ee
where $|W|$ is the order of the Weyl group.%
\footnote{In this paper, we adopt a slightly different normalization of $Z_\text{1-loop}$ than in \cite{Benini:2013nda} such that $Z_{T^2}$ is given by a residue of $Z_\text{1-loop}$ without any additional multiplicative factor.\label{foot: normalization}}
This prescription was found by carefully performing the localization procedure. As discussed in \cite{Benini:2013nda}, the formula generalizes easily to groups with disconnected components.\footnote{In \cite{Gadde:2013dda}, an alternative prescription for $\cN{=}(2,2)$ theories of general rank was given. In all examples studied, the two prescriptions lead to the same results. Note also that the formula up to the choice of cycle $C$ was already given in \cite{Grassi:2007va}, where $C$ was declared to be a cycle that reproduces the Euler number. Also, the formula was derived mathematically in \cite{Gorbounov:2006gs} for complete intersections in products of projective spaces.}

For theories whose gauge group has general rank, we will find the formula
\be
\label{general-schematic}
Z_{T^2} = \frac1{|W|} \sum_{u_* \,\in\, \fM_\text{sing}^*} \JKres_{u=u_*}\!\big( \sQ(u_*), \eta \big) \;\; Z_\text{1-loop} \;,
\ee
where $\JKres(\sQ(u_*),\eta)$ is a residue operation called the \emph{Jeffrey-Kirwan residue}, formulated mathematically in \cite{JeffreyKirwan} by Jeffrey and Kirwan and motivated by a physical discussion by Witten  \cite{Witten:1992xu}.
Here, $\sQ(u_*)$ is the set of charges of the fields responsible for the pole of $Z_\text{1-loop}$ at $u_*$.
In addition this residue operation depends, at each pole $u_*$, on a covector $\eta\in \fh^*$ where $\fh$ is the Cartan subalgebra of the gauge group $G$. Although each of the terms in (\ref{general-schematic}) depends on a choice of $\eta$, the sum does not. As we will see, this corresponds to the fact that the elliptic genus is the same as computed in different phases of a two-dimensional theory.
The formula \eqref{general-schematic}  reproduces the simpler formula \eqref{rank1-schematic} when $G$ has rank 1.

We will present various illustrative examples: Abelian theories that---in their geometric phase---realize Calabi-Yau manifolds as complete intersections in projective spaces, non-Abelian theories that realize complete or incomplete intersections in Grassmannians, and determinantal varieties.
We will also study the dualities involving $U(k)$ and $SU(k)$ gauge theories, some of which are massive and some conformal. Many of these examples were already considered in \cite{Gadde:2013dda}.

In the rest of the paper, we will describe and derive the formula \eqref{general-schematic} in more detail.
We start in section \ref{sec: ell} by setting up the notation, presenting the formula \eqref{general-schematic}, and providing an explanation of the Jeffrey-Kirwan residue operation.
Then in section \ref{sec: derivation}, we derive the formula by localizing the field theory path integral (this section is technical and could be skipped at a first reading).
In section \ref{sec: examples}, we present a few illustrative examples, showing how the formula can be actually used. Further details on our notations are in the appendices, reproduced from \cite{Benini:2013nda} for self-containedness.

\section{Elliptic genera}
\label{sec: ell}

We start by defining the objects of interest, \ie{}, the elliptic genera of two-dimensional theories with $\cN{=}(2,2)$ and $\cN{=}(0,2)$ supersymmetry, and then give a residue formula for them. We adopt the same notation as in \cite{Benini:2013nda}, where the special case of theories with rank-one gauge groups was studied.

\subsection{Theories with $\cN{=}(2,2)$ supersymmetry}
\label{sec: 2,2}

Consider a two-dimensional theory with $\cN{=}(2,2)$ supersymmetry, a flavor symmetry group $K$ (with Cartan generators $K_a$) and a left-moving $U(1)$ R-symmetry $J$ (which is discrete if the theory is not conformal).
Its elliptic genus is defined as
\be
\label{def genus 2,2}
Z_{T^2}(\tau, z, u) =
\Tr_\text{RR} \, (-1)^F q^{H_L} \bar q^{H_R} y^J \prod\nolimits_a x_a^{K_a} \;.
\ee
The trace is taken in the RR sector, \ie{}, we give the fermions periodic boundary conditions. Then $F$ is the fermion number, the parameter
\be
\label{def q}
q = e^{2\pi i \tau}
\ee
specifies the complex structure of a torus $w \sim w+1 \sim w + \tau$, and we write $\tau = \tau_1 + i\tau_2$. $H_L$ and $H_R$ are the left- and right-moving Hamiltonians respectively, defined in Euclidean signature in terms of Hamiltonian and momentum as $2H_L = H + iP$, $2H_R = H - iP$. Since $q^{H_L} \bar q^{H_R} = \exp( - 2\pi \tau_2 H - 2\pi \tau_1 P)$, the trace can be represented by a path integral on a torus of complex structure $\tau$. In a superconformal theory, the operators $H_L, H_R, J$ equal the zero-mode generators $L_0, \bar L_0, J_0$ of the superconformal algebra.%
\footnote{When not uniquely fixed, \eg{} by the superpotential, the superconformal R-symmetries can be determined through the $c$-extremization principle of \cite{Benini:2012cz, Benini:2013cda}.}
We also define the parameters
\be
\label{def y and x}
y = e^{2\pi i z} \;,\qquad\qquad\qquad x_a = e^{2\pi i u_a} \;.
\ee
For a charge vector $\rho^a$, we define
\be
x^\rho = \prod\nolimits_a x_a{}^{\rho^a} = e^{2\pi i \rho^a u_a} \;.
\ee
We often write $\rho(u)=\rho^a u_a$, considering $\rho\in \fk^*$ and $u\in \fk$, where $\fk$ is the Cartan subalgebra of the flavor symmetry group $K$.

In the path integral formulation the effect of $y^J$ and $x_a^{K_a}$ is to turn on flat background gauge fields $A^\text{R}$ and $A^\text{flavor}$ on the torus, coupled to the R-symmetry and flavor symmetry currents respectively, with
\be
\label{def z and u}
z = \oint_t A^\text{R} - \tau \oint_s A^\text{R} \;,\qquad\qquad u_a =\oint_t A^\text{$a$-th flavor} - \tau \oint_s A^\text{$a$-th flavor} \;,
\ee
where $t,s$ are the temporal and spatial cycles.%
\footnote{Choosing a constant connection $A^\text{R}_\mu$, we have $z = (-2i \tau_2)\, A^\text{R}_{\bar w}$ and similarly for the flavor holonomies.}
This is equivalent to specifying non-trivial boundary conditions twisted by the R- and flavor charges, along both the spatial and temporal cycles.%
\footnote{To be precise, in the Hamiltonian definition as written in (\ref{def genus 2,2}), the fields are periodic along the spatial cycle and twisted by complex parameters $y,x_a$ along the temporal cycle.  Instead, we can also take the trace in the sector where the fields are twisted along the spatial cycle by $\oint_s A$, and the chemical potential inside the trace only comes from $\oint_t A$. This matches more directly with the path integral definition where the fields are twisted along both the spatial and temporal cycle by phases $e^{2\pi i \oint_s A}$, $e^{2\pi i \oint_t A}$ respectively. By holomorphy of the result in $z,u_a$, these quantities all coincide.}
When the R-symmetry is discrete, $z$ is only allowed to take certain discrete values.

The elliptic genus when $u_a\neq 0$ is sometimes called the equivariant elliptic genus.
Setting $z=u_a=0$, the elliptic genus reduces to the Witten index, and in particular when the 2d theory has a low energy description as a non-linear sigma model (NLSM), it gives the Euler number of the target manifold. The $q\to0$ limit of the elliptic genus is called the $\chi_y$ genus.

Let us summarize some properties of the $\cN{=}(2,2)$ elliptic genus. Since the spectrum of the Ramond sector is invariant under change conjugation:
\be
Z_{T^2}(\tau,z,u_a) = Z_{T^2}(\tau,-z,-u_a) \;.
\ee
When the R-symmetry is non-anomalous and the theory flows to an IR fixed point, the modular transformations of the elliptic genus are:
\be
\label{modular transform 2,2}
Z \Big( \frac{a\tau + b}{c\tau + d} \,,\, \frac z{c\tau + d} \,,\, \frac{u_a}{c\tau + d} \Big) = \exp\Big[ \frac{\pi i c}{c\tau + d} \Big( -2\cA^a_L u_a z + \frac{c_L}3 z^2 \Big) \Big] \, Z(\tau,z,u)
\ee
with $\smat{a & b \\ c & d} \in SL(2,\bZ)$.
Here $c_L$ is the IR central charge, proportional to the 't~Hooft anomaly of $J$, while $\cA^a_L$ is the t'~Hooft anomaly between $J$ and $K^a$:
\be
c_L = -3 \sum_\text{fermions} \gamma_3 J^2 \;,\qquad\qquad \cA^a_L = \sum_\text{fermions} \gamma_3 J K_a \;.
\ee

In this paper we study gauged linear sigma models (GLSMs), more precisely gauge theories of vector multiplets with matter represented by chiral and twisted chiral multiplets, possibly with superpotential and twisted superpotential interactions. A description of these theories is in \cite{Witten:1993yc}, and we give our conventions in appendix \ref{app: susy}.

We compute the elliptic genus with supersymmetric localization. The BPS configurations relevant for the computation have all bosonic fields set to zero, except for a flat gauge field along the Cartan subalgebra of the gauge and flavor group. Let us parametrize such flat connections by $u$, taking values in the complexified Cartan subalgebra of the gauge and flavor group, as we did in (\ref{def z and u}) for the flavor group alone.
We need the one-loop determinants of quadratic fluctuations around these backgrounds for vector, chiral and twisted chiral multiplets. They have been computed in \cite{Witten:1993jg, Gadde:2013wq, Gadde:2013dda, Benini:2013nda} in a regularization scheme that matches the Hamiltonian computation.

The contribution of a \emph{chiral multiplet} $\Phi$ with vector-like R-charge%
\footnote{The definition (\ref{def genus 2,2}) contains the left-moving R-charge $J$. A chiral multiplet of vector-like R-charge $R$ (and assigning vanishing axial R-charge) has $J = \frac R2$.}
$R$ and transforming in a representation $\fR$ of the gauge and flavor group is:
\be
\label{1-loop chiral}
Z_{\Phi,\fR}(\tau,z,u) = \prod_{\rho \,\in\, \fR} \frac{\theta_1(q, y^{R/2-1} x^\rho)}{\theta_1(q, y^{R/2} x^\rho)} \;.
\ee
The product is over the weights $\rho$ of the representation $\fR$, and $x^\rho \equiv e^{2\pi i \rho(u)}$. Here and in the following we will use interchangeably $\tau,z,u$ and $q,y,x$ using the relations (\ref{def q}) and (\ref{def y and x}).
The function $\theta_1(q,y)$, which we also denote as $\theta_1(\tau|z)$, is a Jacobi theta function and our convention is given in appendix \ref{app: theta}. Notice that if we have two chiral multiplets $\Phi_{1,2}$ in conjugate representations and with R-charges $R_1 + R_2= 2$, then $Z_{\Phi_1,\fR} Z_{\Phi_2, \bar\fR} = 1$ as the two can be given a superpotential mass term and integrated out. Similarly a neutral chiral $\Phi$ with R-charge $R=1$ has $Z_\Phi = -1$.%
\footnote{The minus sign simply follows from a choice of convention for the fermion number. One could choose to include a minus sign in (\ref{1-loop chiral}) instead.}

The contribution of a \emph{vector multiplet} $V$ with gauge group $G$ consists of two parts---the Cartan part with the zero-modes removed and the off-diagonal part (see footnote \ref{foot: normalization}):
\be
\label{1-loop vector}
Z_{V,G}(\tau,z,u) = \bigg( \frac{2\pi\eta(q)^3}{\theta_1(q,y^{-1})} \bigg)^{\rank G} \; \prod_{\alpha \,\in\, G} \frac{\theta_1(q, x^\alpha)}{\theta_1(q, y^{-1} x^\alpha)} \; \prod_{a=1}^{\rank G} \dd u_a\;.
\ee
The product is over the roots $\alpha$ of the gauge group. Then $\eta(q)$ is the Dedekind eta function, with $2\pi \eta(q)^3 = \theta_1'(\tau|0)$ where the derivative is with respect to $z$. Notice that the off-diagonal components give the same contribution as that of twisted chiral multiplets with axial R-charge $2$ and vanishing vector-like R-charge.

Finally, the contribution of a \emph{twisted chiral multiplet} $\Sigma$ with axial R-charge $R_A$ is
\be
Z_\Sigma(\tau,z) = \frac{\theta_1(q, y^{-R_A/2 + 1})}{\theta_1(q, y^{-R_A/2})} \;.
\ee

Notice that all one-loop determinants are meromorphic functions of their arguments and transform under modular transformations according to equation (\ref{modular transform 2,2}).

\subsection{Theories with $\cN{=}(0,2)$ supersymmetry}
\label{sec: 0,2}

Consider a two-dimensional theory with $\cN{=}(0,2)$ supersymmetry and a flavor symmetry group $K$.  The equivariant elliptic genus is defined as
\be
Z_{T^2}(\tau,u) = \Tr_\text{R} (-1)^F q^{H_L} \bar q^{H_R} \prod\nolimits_a x_a^{K_a} \;.
\ee
Again, $q = e^{2\pi i \tau}$ and $x_a = e^{2\pi i u_a}$.
If the theory has a low-energy description as a non-linear sigma model with target a holomorphic vector bundle over a compact complex manifold, as in the models in \cite{Distler:1993mk}, the elliptic genus encodes the Euler number of the vector bundle, see \eg{} \cite{Kawai:1994np}.

Also in this class we study GLSMs, \ie{} gauge theories of vector multiplets with matter in chiral and Fermi multiplets; further interactions are described by potential terms $J_a(\Phi)$ and $E^a(\Phi)$, holomorphic functions of the chiral multiplets and in number equal to the Fermi multiplets (see again \cite{Witten:1993yc} and our appendix \ref{app: susy} for a description of these theories).

The contribution of a \emph{chiral multiplet} $\Phi$ transforming in a representation $\fR$ of the gauge and flavor group is
\be
\label{1-loop chiral 0,2}
Z_{\Phi,\fR}(\tau,u) = \prod_{\rho\,\in\,\fR} i \frac{
\eta(q)}{\theta_1(q, x^\rho)} \;.
\ee
The contribution of a \emph{Fermi multiplet} $\Lambda$ in a representation $\fR$ is
\be
\label{1-loop Fermi}
Z_{\Lambda,\fR}(\tau,u) = \prod_{\rho\,\in\,\fR} i \frac{\theta_1(q,x^\rho)}{
\eta(q)} \;.
\ee
Note that their $q$-expansion can start with a nontrivial power $q^E$, where $E$ is the Casimir energy of the multiplet. Notice also that the product of the determinants of a chiral and a Fermi multiplet in conjugate representations is 1, as they can be given a supersymmetric mass and be integrated out. Moreover suppose we have symmetry group $U(1)_R \times G$: the product of the determinants of a chiral multiplet with charge $\frac R2$, and of a Fermi multiplet with charge $\frac R2-1$, both in representation $\fR$, equals---up to a sign---the determinant (\ref{1-loop chiral}) of an $\cN{=}(2,2)$ chiral multiplet of R-charge $R$.

The contribution of a \emph{vector multiplet} $V$ with gauge group $G$ (with the zero-modes of the Cartan generators removed) is%
\footnote{It was noticed in \cite{Gadde:2013sca} that the non-Abelian vector multiplet determinant serves as the natural measure for the orthogonality of affine characters.}
\be
\label{1-loop vector 0,2}
Z_{V,G}(\tau,u) = \bigg( \frac{2\pi \eta(q)^2}{i} \bigg)^{\rank G} \; \prod_{\alpha \,\in\, G} i \frac{\theta_1(q, x^\alpha)}{\eta(q)} \; \prod_{a=1}^{\rank G} \dd u_a \;.
\ee
Notice that the determinant of an off-diagonal vector multiplet is exactly equal to that of a Fermi multiplet, since in two dimensions the gauge field is non-dynamical and thus the two contain the same degrees of freedom. Moreover, in the case with $U(1)_R \times G$ symmetry, the product of the determinants of a vector (or Fermi) multiplet of charge $0$ and of a chiral multiplet of charge $-1$, reproduces the determinant  (\ref{1-loop vector}) of an $\cN{=}(2,2)$ vector multiplet.

All one-loop determinants are meromorphic functions, and have the following modular transformation properties:
\be
\label{modular transform 0,2}
Z \Big( \frac{a\tau + b}{c\tau + d} \,,\, \frac{u_a}{c\tau + d} \Big) = \epsilon(a,b,c,d)^{c_R-c_L} \; \exp\Big[ - \frac{\pi i c}{c\tau + d} \, \cA^{ab} u_a u_b \Big] \, Z(\tau,u_a)
\ee
with $\smat{a & b \\ c & d} \in SL(2,\bZ)$. The multiplier system $\epsilon(a,b,c,d)$ is a phase, independent of $u_a$, universally defined by
\be
\frac{\eta\big( \frac{a\tau+b}{c\tau + d} \big)}{\theta_1 \big(\frac{a\tau+b}{c\tau+d} \,\big|\, \frac{u}{c\tau+d})} = \epsilon(a,b,c,d) \; e^{-\frac{i\pi c}{c\tau+d} z^2} \; \frac{\eta(\tau)}{\theta_1(\tau|u)} \;.
\ee
It is through $\epsilon$ that the gravitational anomaly shows up.
Finally $\cA^{ab}$ are the flavor 't~Hooft anomalies:
\be
\cA^{ab} = \sum_\text{fermions} \gamma_3 K_a K_b \;.
\ee

\subsection{The formula}
\label{sec: formula}

Let us now present our formula for the elliptic genus of a two-dimensional gauge theory. First of all we construct the one-loop determinant $Z_\text{1-loop}$ with zero-modes removed; such an object is naturally a meromorphic $(r,0)$-form, where $r = \rank(G)$. As described in section \ref{sec: 2,2}, for an $\cN{=}(2,2)$ theory with gauge group $G$, flavor group $K$, and chiral multiplets $\Phi_s$ in representation $\fR_s$ of $G$, with R-charge $R_s$ and weight $K_s$ under the flavor group, from \eqref{1-loop chiral} and \eqref{1-loop vector} we get
\begin{multline}
\label{one-loop 2,2}
Z_\text{1-loop}(\tau,z,u,\xi)= \\
=\bigg( \frac{2\pi\eta(q)^3}{\theta_1(q,y^{-1})} \bigg)^{r} \prod_{\alpha \,\in\, G} \frac{\theta_1(q, x^\alpha)}{\theta_1(q, y^{-1} x^\alpha)}\prod_s \prod_{\rho \,\in\, \fR_s} \frac{\theta_1(q, y^{R_s/2-1} x^\rho e^{2\pi i K_s(\xi)})}{\theta_1(q, y^{R_s/2} x^\rho e^{2\pi i K_s(\xi)})} \, \dd u_1 \cdots \dd u_r \;.
\end{multline}
Notice that the only difference between $u$ and $\xi$ is that $u$ will be integrated over. We will sometimes keep $\xi$ implicit in the following formul\ae.
In the same way, the one-loop determinant $Z_\text{1-loop}(\tau,u,\xi)$ of an $\cN{=}(0,2)$ gauge theory with gauge group $G$, flavor group $K$, and chiral and Fermi multiplets, is formed out of the blocks (\ref{1-loop chiral 0,2}), (\ref{1-loop Fermi}) and (\ref{1-loop vector 0,2}).

The meromorphic form $Z_\text{1-loop}$ has poles in $u$, along hyperplanes corresponding to all chiral and off-diagonal vector multiplets in $\cN{=}(2,2)$, and to chiral multiplets in $\cN{=}(0,2)$. For simplicity, we will assume that the non-Abelian part of $G$ is connected and simply-connected; non-simply-connected and disconnected groups can be treated as well, as in \cite{Benini:2013nda}, but they require more care. Let $\fh$ be the Cartan subalgebra of $G$, then the Cartan torus of $G$ can be identified with $\fh/ {\rm Q}^{\vee}$ where ${\rm Q}^{\vee}$ is the coroot lattice. We define
\be
\label{moduli space}
\fM =\fh_\bC /({\rm Q}^{\vee} + \tau {\rm Q}^{\vee}) \;,
\ee
then the moduli space of flat $G$-connections on $T^2$ is $\fM/W$, where $W$ is the Weyl group. Each of the multiplets listed above introduces a singular hyperplane $H_i \subset \fM$. We will use the index $i$ for them, and call $Q_i \in \fh^*$ the weight of the multiplet under the gauge group. For the different types of multiplets we have:
\begin{equation}
\label{Hi}
\begin{array}{rcr@{\ }lc}
\text{vector}_{(2,2)}: & H_i = \Big\{& - z + Q_i(u) &= 0
\pmod{\bZ+ \tau\bZ } \Big\}  \qquad & Q_i = \alpha, \\[.5em]
\text{chiral}_{(2,2)}:  & H_i = \Big\{& \frac{R_i}2 z + Q_i(u) + K_i(\xi) &= 0 \pmod{ \bZ+ \tau\bZ } \Big\}  &Q_i = \rho, \\[.5em]
\text{chiral}_{(0,2)}:  & H_i = \Big\{& Q_i(u) + K_i(\xi) &= 0
\pmod{ \bZ + \tau\bZ } \Big\}  &Q_i = \rho.
\end{array}
\end{equation}
where $Q_i(u)$ is a pairing between $\fh^*$ and $\fh$. Note also that a single $H_i$ can contain multiple parallel disconnected hyperplanes. We denote by $\sQ = \{Q_i\}$ the set of all charge covectors. Then we define
\be
\fM_\text{sing} = \bigcup\nolimits_i H_i
\ee
in $\fM$, and we denote by $\fM_\text{sing}^* \subset \fM_\text{sing}$ the set of isolated points in $\fM$ where at least $r$ linearly independent hyperplanes meet:
\be
\label{M sing star}
\fM_\text{sing}^* = \big\{ u_* \in \fM \,\big|\, \text{at least $r$ linearly independent $H_i$'s meet at } u_* \big\} \;.
\ee
Given $u_*\in \fM_\text{sing}^*$, we denote by $\sQ(u_*)$ the set of charges of the hyperplanes meeting at $u_*$: \be
\sQ(u_*) = \{  Q_i \,\big|\, u_* \in H_i \} \;.
\ee
For a technical reason, we will assume the following condition: \emph{For any $u_*\in\fM_\text{sing}^*$, the set $\sQ(u_*)$ is  contained in a half-space of $\fh^*$}. A hyperplane arrangement with this property at $u_*$ is called \emph{projective} \cite{SzenesVergne}. Notice that if the number of hyperplanes at $u_*$ is exactly $r$, the arrangement is automatically projective.%
\footnote{When the condition is not met, as in the example of section \ref{sec: ex Rodland}, one needs to relax the constraints on R- and flavor charges coming from the superpotential, resolve $u_*$ into multiple singularities which are separately projective, and eventually take a limit where the charges are the desired ones.}
If at every $u_*$ the number of hyperplanes meeting at $u_*$ is exactly $r$, we call the situation \emph{non-degenerate}.

Denote by $\Cone_\text{sing}(\sQ) \subset \fh^*$ the union of the cones generated by all subsets of $\sQ$ with $r-1$ elements. Then each connected component of $\fh^* \setminus \Cone_\text{sing}(\sQ)$ is called a \emph{chamber}. Choose a generic non-zero $\eta\in \fh^*$, \ie{} an $\eta \not\in \Cone_\text{sing}(\sQ)$: such $\eta$ identifies a chamber in $\fh^*$. Under the assumption, the elliptic genus is given by the formula:
\be
\label{main formula}
Z_{T^2}(\tau,z,\xi) = \frac1{|W|} \sum_{u_* \,\in\, \fM_\text{sing}^*} \JKres_{u=u_*}\!\big(\sQ(u_*),\eta \big) \;\; Z_\text{1-loop}(\tau,z,u,\xi)
\ee
where $|W|$ is the order of the Weyl group. Here $\JKres$ is the Jeffrey-Kirwan residue operation, which is explained in detail below. $\JKres$ is locally constant as a function of $\eta$, but it can jump as $\eta$ crosses from one chamber to another. Nonetheless the sum on the right hand side is independent of $\eta$.

Before proceeding, we note that $\eta\in \fh^*$ should not be confused with the Fayet-Iliopoulos term $\xi\in \fh^*$. When dealing with examples in section \ref{sec: examples}, we will see that $\eta$ and $\xi$ have many similar properties; for instance as $\eta$ is varied over the chambers, (\ref{main formula}) produces the elliptic genus in the various phases of the gauge theory. Nonetheless $\eta$ and $\xi$ are different objects, \eg{} because $\xi$ is only allowed for the Abelian part of the gauge group while we need to choose $\eta$ even for non-Abelian gauge groups. Even for Abelian gauge groups, we do not see any reason why we should take $\eta=\xi$.

\subsection{The Jeffrey-Kirwan residue}

\subsubsection{Defining properties}
The Jeffrey-Kirwan residue operation has been introduced in \cite{JeffreyKirwan}; there are several equivalent formulations available in the literature, and we follow \cite{SzenesVergne}. We define the residue at $u_* = 0$; for generic $u_*$ we just shift the coordinates. Consider $n$ hyperplanes meeting at $u=0\in \bC^r$:
\be
H_i = \big\{ u \in \bC^r \,\big|\, Q_i(u)=0 \big\}
\ee
for $i=1,\ldots, n$ and with $Q_i \in (\bR^r)^*$. Here we indicate the set of charges $\sQ(u_*) = \{Q_i\}$ simply by $\sQ_*$: the charges define the hyperplanes $H_i$ and give them an orientation. The set $\sQ_*$ defines a hyperplane arrangement (for further details on hyperplane arrangements see \eg{} \cite{OrlikTerao}). The coefficients defining the hyperplanes are all real, \ie{} we are dealing with a complexified central arrangement. A residue operation is a linear functional on the space of meromorphic $r$-forms that are holomorphic on the complement of the arrangement, such that it annihilates exterior derivatives of  rational $(r-1)$-forms.

Take a  meromorphic $r$-form $\omega$ defined in a neighborhood $U$ of $u=0$, and holomorphic on the complement of $\bigcup_i H_i$. When $n=r$, we can define the residue of $\omega$ at $u=0$ by its integral over $\prod_{i=1}^r \cC_i$, where each $\cC_i$ is a small circle around $H_i$ (and the overall sign depends on the order of the $H_i$'s). This stems from the fact that the homology group $H_r \big( U \setminus \bigcup_{i=1}^r H_i, \bZ \big) = \bZ$, and therefore there is a natural generator defined up to a sign. When $n>r$ however, $H_r \big( U \setminus \bigcup_{i=1}^n H_i,\bZ \big) = \bZ^{c_{n,r}}$ with $c_{n,r}>1$, and it is imperative to specify the precise cycle to choose.

For a projective arrangement and given an $\eta\in (\bR^r)^*$, the Jeffrey-Kirwan residue is the linear functional defined by the conditions:
\be
\label{SVcondition}
\JKres_{u=0}(\sQ_*,\eta) \, \frac{\dd Q_{j_1}(u)}{Q_{j_1}(u)} \wedge \cdots \wedge \frac{\dd Q_{j_r}(u)}{Q_{j_r}(u)} =
\begin{cases}
\sign \det (Q_{j_1} \dots Q_{j_r}) & \text{if } \eta\in \Cone(Q_{j_1} \ldots Q_{j_r}) \\
0 & \text{otherwise}
\end{cases}
\ee
where $\Cone$ denotes the cone spanned by the vectors in the argument. We can rewrite it as
\be
\label{SVcondition2}
\JKres_{u=0}(\sQ_*,\eta) \,  \frac{\dd u_1 \wedge \dots \wedge \dd u_r}{Q_{j_1}(u) \cdots Q_{j_r}(u)} =
\begin{cases}
\dfrac1{|\det (Q_{j_1} \dots Q_{j_r})|} & \text{if } \eta\in \Cone(Q_{j_1} \ldots Q_{j_r}) \\[1em]
0 & \text{otherwise}
\end{cases}
\ee
after choosing coordinates $u_a$ on $\fh$.
The definition \eqref{SVcondition}-\eqref{SVcondition2} is in general vastly over-determined since there are many relations between the forms $\bigwedge_{\alpha=1}^r \dd Q_{j_\alpha}/Q_{j_\alpha}$, but it has been proven in \cite{BrionVergne} that (\ref{SVcondition}) is consistent.%
\footnote{The definition of the JK residue in \cite{BrionVergne} depends on both a covector $\eta \in \fh^*$ and a vector $\delta \in \fh$, and does not require the arrangement to be projective. If, however, the arrangement is projective one can naturally choose a vector $\delta$ that has positive pairing with all covectors in $\sQ_*$. Any such choice leads to the definition in \cite{SzenesVergne} and that we are using here, which only depends on a covector $\eta$.}
In fact the JK residue is given by an integral over an explicit cycle, as we will review below in section \ref{subsec: explicitcycle}.

\subsubsection{The rank-1 case}
Let us first consider the simplest case $r=1$. Applying (\ref{SVcondition2}) we find
\be
\JKres_{u=0}\big(\{q\}, \eta\big) \; \frac{\dd u}u = \begin{cases} \sign(q) & \text{if } \eta q > 0 \;, \\ 0 & \text{if } \eta q < 0 \;. \end{cases}
\ee
Substituting into \eqref{main formula}, we find that the elliptic genus in the rank-1 case is given by
\be
Z_{T^2} = \frac1{|W|} \sum_{u_+ \,\in\, \fM_\text{sing}^+} \frac1{2\pi i} \oint_{u=u_+} Z_\text{1-loop}
= - \frac1{|W|} \sum_{u_- \,\in\, \fM_\text{sing}^-} \frac1{2\pi i} \oint_{u=u_-} Z_\text{1-loop}\label{rank-1 formula}
\ee
by choosing $\eta = 1$ and $\eta = -1$ respectively. This precisely reproduces the formula \eqref{rank1-schematic} originally found in  \cite{Benini:2013nda}.

\subsubsection{Constructive definition}
\label{subsec: explicitcycle}

A constructive definition of the JK residue has been given in \cite{SzenesVergne}:
\be
\label{JK formula}
\JKres_{u=0}(\sQ_*,\eta) = \sum_{F \,\in\, \cF\cL^+(\sQ_*,\eta)} \hspace{-.2cm} \nu(F) \, \Res_F \;.
\ee
To understand the formula we need some more definitions. First, let $\Sigma\sQ_*$ be the set of elements of $\fh^*$ obtained by partial sums of elements of $\sQ_*$:
\be
\Sigma\sQ_* = \Big\{ {\ts \sum_{i \in \pi} Q_i} \,\Big|\, \pi \subset \{1,\dots,n\} \Big\} \;.
\ee
We impose a stronger regularity condition $\eta \not\in \Cone_\text{sing}(\Sigma\sQ_*)$, \ie{} that $\eta$ does not belong to any hyperplane generated by elements of $\Sigma\sQ_*$ (this implies the weaker $\eta \not\in \Cone_\text{sing}(\sQ_*)$). In fact $\Cone_\text{sing}(\Sigma\sQ_*)$ divides each chamber into sub-chambers, but (\ref{JK formula}) will jump only when $\eta$ moves from one chamber to another.

Then let $\cF\cL(\sQ_*)$ be the finite set of flags
\be
F = \big[ F_0 = \{0\} \subset F_1 \subset \cdots \subset F_r = \fh^* \big] \;,\qquad \dim F_j = j \;,
\ee
such that $\sQ_*$ contains a basis of $F_j$ for each $j=1, \dots, r$. Let the basis of $F_j$ be given by the first $j$ elements of the ordered set $\fB(F) = \{Q_{j_1}, \dots, Q_{j_r}\}$. To each flag $F \in \cF\cL(\sQ_*)$ we associate a linear functional $\Res_F$, called the \emph{iterated residue}, which is simply the residue computed in the basis $\fB(F)$: let $\tilde u_\alpha = Q_{j_\alpha}(u)$ and $\omega = \tilde\omega_{1\dots r} \,d\tilde u_1 \wedge \dots \wedge d\tilde u_r$, then
\be
\Res_F \, \omega = \Res_{\tilde u_r \,=\, 0} \;\cdots\; \Res_{\tilde u_1 \,=\, 0} \, \tilde\omega_{1 \dots r} \label{iterated residue}
\ee
where at each step the other variables are kept constant and generic. The iterated residue only depends on the flag $F$, not on the basis used to compute it, and indeed it corresponds to integrating $\omega$ on a specific cycle \cite{SzenesVergne}.

Third, for each flag $F$ we also introduce the vectors $\kappa_j^F$ that are sums of elements of $\sQ_*$:
\be
\kappa_j^F = \sum\nolimits_{Q_i \,\in\, F_j} Q_i \qquad\text{ for } j = 1,\dots, r \;.
\ee
The number $\nu(F)$ is
\be
\nu(F) = \sign \det (\kappa^F_1 \ldots \kappa^F_r) \;,
\ee
\ie{} it equals $1$ or $-1$ depending on whether the ordered basis $(\kappa_1^F, \dots, \kappa_r^F)$ of $\fh^*$ is positively or negatively oriented, and $\nu(F) = 0$ if the set $\{\kappa_j^F\}$ is linearly dependent.

Finally, consider the closed cone $\fs^+(F,\sQ_*) = \sum_{j=1}^r \bR_{\geq 0} \kappa_j^F$ generated by the elements $\{\kappa_j^F\}$. We denote by $\cF\cL^+(\sQ_*,\eta)$ the set of flags $F$ such that $\eta \in \fs^+(F,\sQ_*)$. Notice that from the stronger regularity condition on $\eta$ it follows that for every flag $F \in \cF\cL^+(\sQ_*,\eta)$, $\nu(F) = \pm 1$.

It is easy to check that when $\sQ_*$ is a set of $r$ linearly independent covectors $\{Q_1, \dots, Q_r\}$, (\ref{JK formula}) agrees with (\ref{SVcondition}). If $\eta \not\in \Cone(Q_1\dots Q_r)$ then $\cF\cL^+(\sQ_*,\eta)$ is empty. Otherwise, the chamber $\Cone(Q_1\dots Q_r)$ is cut by $\Cone_\text{sing}(\Sigma\sQ_*)$ into $r!$ sub-chambers and $\eta$ belongs to one of them. Let this be the cone generated by $\{\kappa^F_j\}$ for the flag generated by $\fB(F) = \{Q_{\pi(1)}, \dots, Q_{\pi(r)}\}$, where $\pi$ is a permutation of $\{1,\dots,n\}$. Such a flag is the only one in $\cF\cL^+(\sQ_*,\eta)$, moreover $\nu(F) = \sign(\pi) \, \sign\det(Q_1 \dots Q_r)$. Applying (\ref{JK formula}) to the form $\omega = \frac{\dd Q_1(u)}{Q_1(u)} \wedge \dots \wedge \frac{\dd Q_r(u)}{Q_r(u)}$ we get (\ref{SVcondition}).

\section{Derivation via localization}
\label{sec: derivation}

The aim of this section is to derive the formula \eqref{main formula} by a localization computation.
Those who are more interested in how the formula is used, and those who trust the authors, can proceed directly to section \ref{sec: examples} where many illustrative examples are discussed.

Before getting into the details, we would like to spend a few paragraphs to motivate why our derivation is going to be rather delicate and subtle. A schematic way to explain the supersymmetric localization often goes as follows. We consider an integral over a supermanifold $\cM$, with an action of a fermionic symmetry $\cQ$, acting on bosonic and fermionic coordinates $x$ and $\eta$:
\be
Z = \int_\cM \dd x\, \dd \eta\, e^{-S} \;.
\ee
We can add an exact term $-\e^{-2} \big( \sum_\xi |\cQ\xi|^2 + \text{fermionic} \big) $ to the action---where $\xi$ are fermionic variables in the system---without changing the integral. Then we have
\be
\label{starting point}
Z = Z(\e) = \int_\cM \dd x\, \dd \eta\, e^{- S - \e^{-2} ( \sum_\xi |\cQ\xi|^2 + \text{fermionic})} \;.
\ee
We take $\e\to0$, thus localizing the integral on the BPS subspace
\be
\cM_\text{BPS} = \big\{ x \in \cM \,\big|\, \cQ\xi= 0 \text{ for all } \xi \big\} \;.
\ee
We end up with the formula
\be
\label{naive localization formula}
Z = \int_{\cM_\text{BPS}} \omega
\ee
where $\omega$ is the differential form resulting from the fermionic and bosonic Gaussian integral around $\cM_\text{BPS} \subset \cM$.

In our situation the naive localization formula \eqref{naive localization formula} does not make sense, because $\omega$ is generically zero on $\cM_\text{BPS}$ due to fermionic zero-modes coming from the gaugini, and because the part without fermionic zero-modes diverges on a subset $\cM_\text{sing} \subset \cM_\text{BPS}$. This signifies the break-down of the assumption that the contribution from an infinitesimal neighborhood of $\cM_\text{BPS}$ within $\cM$ is well under control even in the limit $\e \to 0$.
Eventually, in our particular case, we find a formula of the form
\be
\label{result}
Z = \int_C \tilde\omega
\ee
where $C$ is a middle-dimensional cycle in $\cM_\text{BPS}$, and $\tilde\omega$ is what results from the Gaussian integral around $\cM_\text{BPS} \subset \cM$ if we drop the fermionic zero-modes that make $\omega$ vanish. In fact the combined one-loop factor $Z_\text{1-loop}$---\eg{} in \eqref{one-loop 2,2}---appearing in our main formula \eqref{main formula} is such $\tilde\omega$, which differs from the full Gaussian integral over all modes. It is relatively easy to obtain $Z_\text{1-loop} = \tilde\omega$, but this schematic derivation is too crude to determine $C$.  It is in fact \emph{too late} if we reached the stage \eqref{naive localization formula}.

Rather, we need to take the limit $\e \to 0$ in \eqref{starting point} carefully, \eg{} by estimating how big a tubular neighborhood needs to be kept around each point of $\cM_\text{BPS}$, so that the apparent divergence in $\omega$ does not affect the limiting procedure. Therefore to obtain \eqref{result} we need to: i) split $\cM_\text{BPS}$ into regions; ii) perform various estimates and take the limit $\e\to0$ carefully in each region; iii) combine the contributions from the various regions. This is what we are going to do in this section. Before proceeding, the reader is advised to go through the analysis of the rank-one case (presented in section 3 of our previous paper \cite{Benini:2013nda}) because it is much simpler and yet it contains the physical idea.

\tocless\subsection{The quantity to compute}

The part of the localization procedure sketched in this subsection---to get the quantity of interest (\ref{D reinstated})-(\ref{fuD})---is essentially the same as in sections 3.2.1 and 3.2.2 of \cite{Benini:2013nda}, and we refer there for more details.

We will denote the gauge coupling by $\e$, so that there is a factor $1/\e^2$ in front of the gauge kinetic terms. We also put a factor $1/\g^2$ in front of the kinetic terms for chiral (and possibly Fermi) multiplets. These terms are $\cQ$-exact, so we can perform the localization by sending $\e$ and $\g$ to zero.

Naively the locus on which the path integral localizes is the space of flat connections on $T^2$, parameterized by $u\in \fM$ (\ref{moduli space}) up to the identifications by the Weyl group. After properly taking care of the fermionic zero-modes of the left-moving gaugini---as we do in (\ref{fuD})---the total one-loop factor $Z_\text{1-loop}$ around a given $u$ still diverges at the hyperplanes $H_i$ \eqref{Hi} due to scalar zero-modes in chiral and $\cN{=}(2,2)$ vector multiplets.

Let us denote the union of the singular hyperplanes by $\fM_\text{sing}=\bigcup_i H_i$.  To cope with the divergence, we first fix a very small but finite $\e$. Then we remove an $\varepsilon$-neighborhood
$\Deps \fM_\text{sing}$ of the dangerous region from the integration domain $\fM$, perform the integral, and take $\varepsilon \to 0$. Eventually we take the limit $\e \to 0$. Keeping $\e$ finite during the process removes the divergences on $\fM_\text{sing}$ and guarantees the correctness of the $\varepsilon\to0$ limit. Alternatively one can take a scaling limit $\e,\varepsilon \to 0$ with $\varepsilon < \e^\#$ for a large enough power (see \cite{Benini:2013nda}). Either way, we will indicate such a limit by $\lim_{\e,\varepsilon\to0}$. At this stage it is convenient to reinstate the vector multiplet auxiliary field $D\in \fh$. The quantity to compute is
\be
\label{D reinstated}
Z_{T^2} = \frac1{|W|} \, \lim_{\e,\varepsilon\to0} \int_{\fh} \dd^rD \int_{\fM\setminus\Delta_\varepsilon} \hspace{-1em} \dd^{2r}u \; f_\e(u,D) \, \exp\Big[ - \frac1{2\e^2} D^2 - i\xi(D) \Big] \;,
\ee
where $f_\e(u,D)$ is the result of the path integral over all modes except for the flat connection zero-mode $u$ and $D$. We allowed a Fayet-Iliopoulos term $\xi \in \fh^*$.

The function $f_\e(u,D)$ has a smooth limit as $\e\to0$, so we can take the limit immediately (we cannot do the same with the exponential, as we explained). In the $\cN{=}(2,2)$ case one finds
\bea
\label{fuD}
f_\e(u,D) &\xrightarrow[\e\to0]{} \int \prod_c \dd\lambda_{c,0} \, \dd\bar\lambda_{c,0}\, \bigg\langle \prod_{a,b} \int \dd^2x\, \lambda_a \sum_i Q_i^a \psi_i \phi_i \int \dd^2x\, \bar\lambda_b \sum_j Q^b_j \bar\psi_j \bar\phi_j \bigg\rangle_\text{free} \\[.1cm]
&\qquad = \det\big[ h^{ab}(\tau,z,u,D) \big] \, g(\tau,z,u,D) \;.
\eea
In the first line we took care of the fermionic zero-modes of the left-moving gaugini; $a,b,c$ are gauge indices for the Cartan part, while $i,j$ run over chiral and off-diagonal vector multiplets. Then
\be
h^{ab} = c \sum_i \sum_{m,n \in \bZ} \frac{Q_i^a Q_i^b}{\Big( \big| m+n\tau + \tfrac{R_i}2 z + Q_i( u) \big|^2 + i Q_i( D) \Big) \big( m + n\bar\tau + \tfrac{R_i}2 \bar z + Q_i( \bar u) \big)}
\ee
and $g(\tau,z,u,D)$ is the one-loop factor evaluated at non-zero $D$, whose explicit form was given in \cite{Benini:2013nda}, sections 2.1  and 2.2. In particular $g(\tau,z,u,0) = Z_\text{1-loop}(\tau,z,u)$.
The overall constant $c$ can be fixed by comparing with a single example, as we will do at the very end. In the $\cN{=}(0,2)$ case one finds essentially the same expression%
\footnote{In fact, even in the $\cN{=}(2,2)$ case, we make a  slight redefinition of the auxiliary fields of $\cN{=}(2,2)$ non-Abelian vector multiplets compared to the normal conventions in the literature, so that they nicely decompose into $\cN{=}(0,2)$ multiplets. More details can be found around \eqref{eq:abcde} in appendix \ref{app: susy}.}
(there is no left-moving R-symmetry and $z$, but there are still flavor symmetries and $\xi_\alpha$ that we kept implicit here), with a sum over chiral multiplets only.

We remark that the function $g(\tau,z,u,D)$, that $\eg$ for an $\cN{=}(2,2)$ chiral multiplet reads
\be
\label{one-loop with D}
Z_{\Phi,Q}(\tau,z,u,D) = \prod_{m,n} \frac{\big( m+n\tau+(1- \tfrac R2 ) z - Q(u) \big) \big( m + n \bar\tau + \frac R2 \bar z + Q(\bar u) \big)}{ \big| m + n\tau + \frac R2 z + Q(u) \big|^2 + i Q(D)} \;,
\ee
at generic non-zero $D \in \fh$ does not have any divergence in $u$, so indeed keeping $\e$ finite in (\ref{D reinstated}) removes all singularities.

\tocless\subsection{Expressions in terms of differential forms}

We can cast \eqref{D reinstated} with \eqref{fuD} in a more compact form.
The symmetric matrix function $h^{ab}$ satisfies the following two properties:
\be
\label{with index}
\parfrac{h^{ab}}{\bar u_c} = \parfrac{h^{cb}}{\bar u_a} \;,\qquad\qquad\qquad \parfrac{g}{\bar u_a} = \frac ic\,  h^{ab} D_b g \;,
\ee
as can be seen from an explicit calculation. Let us introduce the $(0,1)$-forms $\nu^b$ on $\fM$ and the Dolbeault operator $\bar\partial$:
\be
\nu^b \,\equiv\, \dd\bar u_a \, h^{ab} \;,\qquad\qquad\qquad \bar\partial = \dd\bar u_a \, \parfrac{}{\bar u_a} \;.
\ee
Regarding $(\nu^a)_{a=1,\ldots,r}$ as a single $\fh_\bC^*$-valued one-form $\nu$ on $\fM$, we can rewrite the relations \eqref{with index} succinctly as
\be
\dbar \nu=0 \;,\qquad\qquad\qquad\qquad \dbar g = \frac ic\, \nu(D)\,g \;.
\ee
The intermediate form \eqref{D reinstated} of the elliptic genus can then be written simply as
\be
Z_{T^2}(\tau,z) = \frac1{|W|} \, \lim_{\e,\varepsilon\to0} \int_\intd{-3pt}{\fh \times (\fM\setminus\Deps)} \hspace{-1cm} \mu
\ee
where
\be
\mu \,\equiv\, g \, \exp\Big[ - \frac1{2\e^2} D^2 - i\xi(D) \Big] \; \dd^ru\wedge \big( \nu(\dd D) \big)^{\wedge r} \;.
\ee
So far we have taken $D$ to be valued in $\fh$. However it will prove convenient to analytically continue the integrand to $D \in \fh_\bC$ and shift the integration contour of $D$ to $\Gamma = \fh+i\delta$, where $\delta\in \fh$ is a constant element that we will choose below. We thus have
\be
\label{to-be-evaluated}
Z_{T^2}(\tau,z) = \frac1{|W|} \, \lim_{\e,\varepsilon\to0} \int_\intd{-3pt}{\Gamma \times (\fM \setminus \Deps)} \hspace{-1cm} \mu \qquad .
\ee
As is clear from (\ref{one-loop with D}), $g(\tau,z,u,D)$ has poles in the complex $D$-plane along the imaginary axes for all fixed $u\in \fM$, and they approach $D=0$ as $u$ approaches $\fM_\text{sing}$. On the other hand these poles are safely at a distance at least of order $\varepsilon$ from the real $D$-axes, as long as we keep $u\in \fM\setminus\Deps$. Therefore the result of (\ref{to-be-evaluated}) is independent of $\delta$ as long as it is sufficiently close to zero, because the integrand does not have any pole on $\fh_\bC \times (\fM\setminus\Deps)$ as we stay sufficiently close to the real $D$-lines.

For reasons that will become clear later, we require that $Q_i(\delta)\neq 0$ for all $Q_i \in \sQ$. We will impose more conditions in section \ref{sec: iterated}.

\tocless\subsection{A helpful identity}

Given a set of charge vectors $\{Q_1, \dots, Q_s\} \subset \fh^*$, define
\begin{multline}
\label{def mu Q's}
\mu_{Q_1, \dots, Q_s} \,\equiv\, \frac{(ic)^s}{(r-s)!} \; g \, \exp\Big[ - \frac1{2\e^2} D^2 - i\xi(D) \Big] \\
\dd^r u \wedge \big( \nu(\dd D) \big)^{\wedge (r-s)} \wedge \frac{\dd Q_1(D)}{Q_1(D)} \wedge \dots \wedge \frac{\dd Q_s(D)}{Q_s(D)} \;.
\end{multline}
This is an $(r,r-s)$-form in $u$-space, in particular $\partial\mu_{Q_1,\dots,Q_s}=0$, and an $s$-form in $D$-space. Note that $\dd Q_a(D) = Q_a(\dd D)$, and $\mu_{Q_1,\dots,Q_s}$ vanishes if $Q_1,\ldots, Q_s$ are linearly dependent. We find:
\be
\label{Fid}
\dd \mu_{Q_0,\dots, Q_s} =  \sum_{i=0}^s (-1)^{s-i} \, \mu_{Q_0, \dots \widehat Q_i \dots, Q_s} \;,
\ee
where $\widehat{\;}$ means omission.

To prove it, first define the $(r-n)$-forms
\be
\omega_{a_1 \dots a_n} \,\equiv\, \frac1{(r-n)!} \, \nu^{b_1} \ldots \nu^{b_{r-n}} \, \epsilon_{b_1 \dots b_{r-n} a_1 \dots a_n} \;,
\ee
where $\epsilon$ is the antisymmetric symbol. They satisfy $\bar\partial \omega_{a_1 \dots a_n} = 0$, and there are two special cases: $\omega = \nu^1 \ldots \nu^r$ and $\omega_{a_1 \dots a_r} = \epsilon_{a_1 \dots a_r}$. Then, using
\be
\epsilon^{a_1 \dots a_{r-n} b_1 \dots b_n} \, \epsilon_{a_1 \dots a_{r-n} c_1 \dots c_n} = (r-n)!\, n!\, \delta^{b_1}_{[c_1} \dots \delta^{b_n}_{c_n]} \;,
\ee
we get:
\be
\bar\partial \big( g \, \omega_{a_1 \dots a_n} \big) = \frac ic\, (-1)^{r-n} g\, n\, D_{[a_1} \omega_{a_2 \dots a_n]} = \frac ic\, (-1)^{r-n} g \sum_{i=1}^n (-1)^{i-1} D_{a_i} \omega_{a_1 \dots \widehat a_i \dots a_n} \;.
\ee
The same equations are valid if we multiply by the exponential in (\ref{def mu Q's}). The form $\mu_{Q_1,\dots,Q_s}$ can be written as
\be
\mu_{Q_1,\dots,Q_s} = (ic)^s \, g \, \exp\Big[ - \frac1{2\e^2} D^2 - i\xi(D) \Big] \; \dd^r u \wedge \omega_{a_1 \dots a_s} \, \frac{Q_1^{a_1} \dots Q_s^{a_s}}{Q_1(D) \ldots Q_s(D)} \, \dd^r D \;.
\ee
Then we just compute $\dd\mu_{Q_0,\dots,Q_s} = \bar\partial \mu_{Q_0,\dots, Q_s}$, thus obtaining \eqref{Fid}.

\tocless\subsection{Cell decomposition of $\fM$}
\label{sec: cells}

The basic idea behind the remaining computation is to apply the identities \eqref{Fid} repeatedly $r$ times to \eqref{to-be-evaluated}, so that we end up with a residue integral. To do that, we first need to construct a suitable cell decomposition of $\fM$.

Define the open $\varepsilon$-neighborhoods
\be
\Deps(H_i) = \big\{ u \in \fM \,\big|\, |Q_i(u) + \cdots| < \varepsilon \big\}
\ee
of the singular hyperplanes $H_i \subset \fM$ (for the $i$-th field), where the dots stand for the constant shifts as described in (\ref{Hi}); we also define their union
\be
\Deps \equiv \bigcup\nolimits_i \Deps(H_i) \;.
\ee
We will study the integral over the closed set $\fM \setminus \Deps$.
The boundary $\partial \Deps$ of the integration domain can be separated into tube regions
\be
S_i \,\equiv\, \partial \Deps \cap \partial \Deps(H_i) \;.
\ee
We give them the natural orientation. We have%
\footnote{On the one hand $\bigcup_i S_i \subset \partial\Deps$. On the other hand, using $\partial (\bigcup_i A_i) \subset \bigcup_i \partial A_i$, we have
$$
{\ts
\bigcup_i S_i = \partial\Deps \cap \bigcup_i \partial\Deps(H_i) \,\supset\, \partial\Deps \cap \partial \big( \bigcup_i \Deps(H_i) \big) = \partial\Deps \;.
}
$$
}
\be
\label{decomposition partial Deps}
\partial\Deps = \bigcup\nolimits_i S_i \;.
\ee
In fact we can show something stronger: the union is quasi-disjoint. Let us introduce
\be
S_{i_1\ldots i_s} \,\equiv\, S_{i_1}\cap \cdots \cap S_{i_s}
\ee
with the natural orientation induced from the natural one on $\partial \Deps(H_{i_1}) \cap \cdots \cap \partial \Deps(H_{i_s})$. They are totally antisymmetric in $i_1,\ldots,i_s$ and
\be
\label{decomposition S}
\partial S_{i_1\ldots i_s} = - \bigcup\nolimits_j S_{i_1\ldots i_s j} \;.
\ee
Indeed $S_{i_1 \ldots i_s}$ is a closed set, then $\partial S_{i_1\ldots i_s} \subset S_{i_1\ldots i_s}$; on the other hand the boundary is where $S_{i_1\ldots i_s}$ meets the other $\Deps(H_j)$, and the sign follows from the natural orientation.
Each manifold $S_{i_1\ldots i_s}$ has real dimension $2r-s$, unless it is empty. Indeed on the one hand the
manifold $\big\{ |Q_{i_1}(u) + \cdots| = \cdots = |Q_{i_s}(u) + \cdots| = \varepsilon\big\}$ has dimension $\geq 2r-s$, if not empty; on the other hand $S_{i_1\ldots i_s}$ is part of the boundary of $S_{i_1\ldots i_{s-1}}$ and proceeding by induction it has dimension $\leq 2r-s$. We conclude that the decompositions in (\ref{decomposition partial Deps}) and (\ref{decomposition S}) are almost disjoint---in the sense that every intersection has dimension lower than the components---and the integral of the union is the sum of the integrals.

The rest of this subsection will be spent in constructing $2r$-dimensional cycles $C_i$ and $(2r-p+1)$-dimensional cycles $C_{j_1\ldots j_p}$, antisymmetric in $j_1$, \ldots, $j_p$, such that the relations
\be
\label{main decomposition}
\fM \setminus \Deps = \bigsqcup_{i} C_i
\ee
and
\be
\label{boundary decomposition}
\partial C_{j_1\ldots j_p}= - S_{j_1\ldots j_p} + \sum\nolimits_i C_{j_1\ldots j_p i}
\ee
hold. Here  the subscripts $i$, $j_{1,\ldots,p}$ are the same ones that label the charge vectors $Q_i$. They are constructed as follows.

First, construct a cell decomposition of $\fM\setminus\Deps$ which is as good as possible.
We call a cell decomposition \emph{good} if a codimension-$k$ cell is at the intersection of $k+1$ codimension-$(k-1)$ cells (\ie{} a codimension-1 cell is at the intersection of two maximal-dimensional cells, a codimension-2 cell is at the intersection of three codimension-1 cells, \etc). Since $\fM \setminus \Deps$ is a manifold with boundary and corners, we cannot construct a good decomposition of it, but we will take one as good as possible. We require the following conditions (i)--(iii).
(i) Each cell is such that its interior is either in the interior of $\fM \setminus \Deps$, or in the interior of exactly one $S_{i_1 \dots i_s}$.
(ii) The cell decomposition is good in the interior of $\fM \setminus \Deps$. To describe the final condition, we note that a neighborhood in $\fM \setminus \Deps$ of an interior point of a corner $S_{i_1\ldots i_s}$ is of the form $\R_+^s \times U$ with $U \subset S_{i_1\ldots i_s}$, and we can think of $U\subset \R^{2r-s}$.
More explicitly, the neighborhood is a domain in $\R^{2r} = \{ (x_{i_1}, \ldots ,x_{i_s},y_{s+1}, \ldots, y_{2r})\}$ defined by $x_{i_1} \geq 0,\ldots, x_{i_s}\geq 0$ and $\vec y \in U$. For $\{j_1,\ldots,j_p\}\subset\{i_1,\ldots, i_s\}$, the corner $S_{j_1\ldots j_p}$ includes $S_{i_1\ldots i_s}$ and a patch of it is identified with the region $x_{j_1}=\ldots=x_{j_p}=0$.
We introduce a cell decomposition of $\R_+^s \times U$ as follows. Given a good cell decomposition of $U$, for $\{j_1,\ldots,j_p\} \subset \{i_1,\ldots, i_s\}$ ($0 \leq p \leq s$) and distinct we define the cells of type $\hat C_{j_1\ldots j_p}^{(2r-k+1)}[i_1,\dots,i_s,U]$ (with $p \leq k-1 \leq 2r-s+p$) by the conditions
\be
0 = x_{j \in \{j_1,\dots,j_p\}} \;,\qquad 0 \leq x_{j \not\in \{j_1, \dots, j_p\}} \;,\qquad \vec y \in (2r-k+1-s+p)\text{-dimensional cell of } U \;.
\ee
The notation is that the cell lives in the neighborhood of an interior point of $S_{i_1,\dots, i_s}$, patch $U$, it has dimension $(2r-k+1)$, and it is the product of a $(2r-k+1-s+p)$-dimensional cell of $U$ and of an $(s-p)$-dimensional quadrant in $\bR_+^s$. The orientation of $\hat C^{(2r-p)}_{j_1,\dots,j_p}[i_1,\dots, i_s,U]$ is the one induced from $S_{i_1,\dots,i_p}$ and we can similarly assign a natural orientation to all the others. This gives antisymmetry in $j_1,\dots,j_p$. The boundary of a cell of type $\hat C_{j_1\ldots j_p}^{(2r-k+1)}[i_1,\dots,i_s,U]$ is the union of  cells of type $\hat C_{j_1\ldots j_p}^{(2r-k)}[i_1,\dots,i_s,U]$ taking the boundary cells in $U$, and of $\bigcup_{j \in \{i_1,\dots,i_s\}\setminus\{j_1,\dots,j_p\}} \hat C_{j_1\ldots j_pj}^{(2r-k)}[i_1,\dots,i_s,U]$ taking the same cell in $U$. We can now describe the condition (iii): a cell touching the interior of $S_{i_1\ldots i_s}$, coincides with a cell of type $\hat C_{j_1\ldots j_p}^{(2r-k+1)}[i_1,\dots,i_s,U]$.

The cells of type $\hat C_{j_1\ldots j_p}^{(2r-k+1)}[i_1,\dots,i_s,U]$ introduced above refer to specific neighborhoods $\R_+^s \times U$ of $S_{i_1,\ldots,i_s}$. In addition we have the cells of type $\hat C^{(2r-k+1)}$, not touching any $S_{\dots}$, from the decomposition of the interior of $\fM\setminus\Deps$. Now, to each $(2r-k+1)$-dimensional cell we assign a set of $k$ (not necessarily distinct) charge vectors $Q_i$, assigned by a function $\fC$ as follows. To the $2r$-dimensional cells of type $\hat C^{(2r)}$ and of type $\hat C^{(2r)}[i_1,\dots, i_s,U]$ we assign a charge vector $Q_i$ randomly, and write $\{Q_i\} = \fC(C)$ where $C$ is a $2r$-dimensional cell. To a cell $C$ of type $\hat C^{(2r-k+1)}$ we assign $k$ charge vectors determined by the $k$ $2r$-dimensional cells of type $\hat C^{(2r)}$ surrounding it (because the decomposition is good): $\{Q_{i_1},\ldots,Q_{i_k}\} = \fC(C)$. To a cell of type $\hat C_{j_1\ldots j_p}^{(2r-k+1)}[i_1,\dots,i_s,U]$ we assign the $k-p$ charge vectors determined by the $k-p$ cells of type $\hat C^{(2r)}_{j_1\ldots j_p}[i_1,\dots,i_s,U]$ surrounding it in the good decomposition of $U$, as well as the vectors $Q_{j_1}, \dots, Q_{j_p}$. This concludes the construction of the cell decomposition.

Now, for $j_1,\dots,j_k$ distinct, we define the $(2r-k+1)$-dimensional domains
\be
C_{j_1\ldots j_k} = \bigcup \bigg\{ \text{cells of type } \hat C^{(2r-k+1)} \text{ and } \hat C^{(2r-k+1)}_{j_1\dots j_p}[i_1\dots i_s,U] \;\Big|\; \fC(\text{cell}) = \{Q_{j_1}, \dots, Q_{j_k} \} \bigg\}
\ee
where  the  union is over all cells in the decomposition, including all $i_1,\dots, i_s$ and all $U$.
By construction, the domains $C_{j_1\ldots j_k}$ satisfy the conditions \eqref{main decomposition} and \eqref{boundary decomposition}.

\tocless\subsection{Cycles in $D$-space}
\label{sec: cyclesinD}

In the following we will also need various integration contours in the complexified $D$-space $\fh_\bC$, besides $\Gamma$. The motivation behind the following definitions will become clear  in section \ref{sec: shift}.

We already defined the contour $\Gamma$ for the $D$-integral in (\ref{to-be-evaluated}):
\be
\Gamma = \big\{ D \in \fh_\bC \,\big|\, \im D = \delta \big\} \;.
\ee
It has the topology of $\bR^r$ and an imaginary shift by a chosen vector $\delta$.

Next we introduce the contours $\Gamma_{i_1 \dots i_p}$: they have the topology of $\bR^{r-p} \times T^p$, circle around $\bigcap_{k=1}^p \big\{ Q_{i_k}(D) = 0 \big\}$ and have an imaginary shift by chosen vectors $\delta_{i_1 \dots i_p}$. For consistency such vectors must satisfy
\be
\label{vectors delta i}
0 = Q_{i_1}(\delta_{i_1 \dots i_p}) = \ldots = Q_{i_p}(\delta_{i_1 \dots i_p}) \;.
\ee
To define the contours, we first construct the loops $\ell_{i_1 \dots i_p} \simeq T^p$ in $\fh_\bC$ which circle around $\bigcap_{k=1}^p \big\{ Q_{i_k}(D) = 0 \big\}$ and stay sufficiently close to the origin. Then
\be
\Gamma_{i_1 \dots i_p} = \big\{ D \in \fh_\bC \,\big|\, \im D = \delta_{i_1 \dots i_p} \;,\quad Q_{i_1}(D) = \ldots = Q_{i_p}(D) = 0 \big\} + \ell_{i_1 \dots i_p} \;.
\ee

Finally we introduce the contours $\Gamma_{i_1 \dots i_p \slj_1 \dots \slj_q}$: they are defined in the same way as $\Gamma_{i_1 \dots i_p}$, but instead of shifting the imaginary part by $\delta_{i_1 \dots i_p}$ we shift it by another vector $\delta'$ with the further constraint $Q_j(\delta') < 0$ for $j \in \{j_1,\dots, j_q\}$. There is always such a $\delta'$ if $Q_{i_1}, \dots, Q_{i_p},Q_{j_1},\dots Q_{j_q}$ are linearly independent. We will not need to give a name to $\delta'$. The orientation is the natural one so that we have, for example,
\be
\label{Gammadefo}
\Gamma = \Gamma_i + \Gamma_{\sli}
\ee
in homology.

\tocless\subsection{Application of Stokes' theorem}
\label{sec: prep}

At this point we can use the identity (\ref{Fid}) and the cell decomposition of the integration domain $\fM\setminus\Deps$ to simplify the integral $\int_{\Gamma \times (\fM\setminus\Deps)} \mu$ somewhat.
We use $\mu=\dd\mu_{Q_i}$ (\ref{Fid}) in each $C_i$.
Then we have
\be
\label{XXX}
\int_\intd{-3pt}{\Gamma\times\fM\setminus\Deps} \hspace{-.5cm} \mu =
\sum_i \int_{\Gamma\times C_i} \dd\mu_{Q_i} = \sum_i \int_{\Gamma\times\partial C_i} \mu_{Q_i}
= - \sum_i \int_{\Gamma\times S_i} \mu_{Q_i} + \sum_i \sum_{j\,(\neq i)} \int_{\Gamma\times C_{ij}} \mu_{Q_i}
\ee
where we used \eqref{main decomposition} and \eqref{boundary decomposition}. The second term in the last expression can be further simplified using \eqref{Fid} and the antisymmetry of $C_{i_1\dots i_p}$:
\be
\sum_i \sum_{j \, (\neq i)} \int_{\Gamma\times C_{ij}} \mu_{Q_i} = \sum_{i<j} \int_{\Gamma\times C_{ij}} \big( \mu_{Q_i}-\mu_{Q_j} \big)
= \sum_{i<j} \int_{\Gamma\times C_{ij}} \dd\mu_{Q_i,Q_j} = \sum_{i<j} \int_{\Gamma\times \partial C_{ij}} \mu_{Q_i,Q_j} \;.
\ee
Plugging it back into \eqref{XXX} and using (\ref{boundary decomposition}) again, we find
\be
\int_\intd{-3pt}{\Gamma\times \fM\setminus\Deps} \hspace{-.5cm} \mu = - \sum_i \int_{\Gamma\times S_i} \mu_{Q_i} -  \sum_{i<j} \int_{\Gamma\times S_{ij}}\mu_{Q_i,Q_j} + \sum_{i<j} \, \sum_{k \, (\neq i,j)} \int_{\Gamma\times C_{ijk}} \mu_{Q_i,Q_j} \;.
\ee
The procedure can be repeated, stopping when we reach the middle-dimensional cohomology in $\fM$, because $\mu_{Q_{i_1},\ldots,Q_{i_p}} = 0$ when $p>r$.
 We obtain:
\be
\label{eqeq}
\int_\intd{-3pt}{\Gamma\times(\fM\setminus\Deps)} \hspace{-1cm} \mu \hspace{.6cm} =
- \sum_i \int_\intd{-3pt}{\Gamma\times S_i} \hspace{-.5cm} \mu_{Q_i} - \sum_{i<j} \int_\intd{-3pt}{\Gamma\times S_{ij}} \hspace{-.5cm} \mu_{Q_i,Q_j} + \ldots - \sum_{i_1<\cdots <i_r} \int_\intd{-3pt}{\Gamma\times S_{i_1\ldots i_r}} \hspace{-1cm} \mu_{Q_{i_1},\ldots,Q_{i_r}} \;.
\ee

Next we show that a similar formula holds when integrating $D$ over the other contours $\Gamma_{i_1\dots i_p}$ of section \ref{sec: cyclesinD}: for instance
\be
\label{eqeq2}
\int_\intd{-3pt}{\Gamma_i \times S_i} \hspace{-.5cm} \mu_{Q_i} = - \sum_j \int_\intd{-3pt}{\Gamma_i \times S_{ij}} \hspace{-.6cm} \mu_{Q_i,Q_j} - \sum_{j<k} \int_\intd{-3pt}{\Gamma_i \times S_{ijk}} \hspace{-.8cm} \mu_{Q_i,Q_j,Q_k} + \ldots - \sum_{j_1 < \ldots < j_{r-1}} \int_\intd{-3pt}{\Gamma_i \times S_{ij_1 \ldots j_{r-1}}} \hspace{-1cm} \mu_{Q_i,Q_{j_1},\ldots,Q_{j_{r-1}}}
\ee
and more generally
\be
\label{eqeq3}
\int_\intd{-3pt}{\Gamma_{i_1 \dots i_p} \times S_{i_1 \dots i_p}} \hspace{-1cm} \mu_{Q_{i_1}, \ldots ,Q_{i_p}} = - \sum_{m=p+1}^r \Bigg[ \sum_{i_{p+1} < \ldots < i_r} \int_\intd{-3pt}{\Gamma_{i_1 \dots i_p} \times S_{i_1 \dots i_m}} \hspace{-1.2cm} \mu_{Q_{i_1}, \ldots, Q_{i_m}} \Bigg] \;.
\ee
The relation \eqref{eqeq} can be thought of as the case $p=0$ of the formula above.
Consider for instance the integral over $\Gamma_i \times S_i$ in (\ref{eqeq2}).
If $Q_{j_1},\ldots,Q_{j_s}$ are linearly independent of $Q_i$, the form $\mu_{Q_{j_1},\ldots,Q_{j_s}}$ has no poles in the region surrounded by $\Gamma_i$ and hence it vanishes when integrated over $\Gamma_i \times S_i$. The identity (\ref{Fid}) then reduces to
\be
\dd\mu_{Q_i,Q_{j_1}, \dots, Q_{j_s}} \,\simeq\, \sum_{k=1}^s (-1)^{s-k} \mu_{Q_i,Q_{j_1},\ldots \widehat Q_{j_k} \ldots,Q_{j_s}}
\ee
when integrated over $\Gamma_i \times S_i$. If instead the vectors are linearly dependent, the formula is trivially true.

Recall that $S_i$ is a manifold with boundary and corners consisting of $S_{ij_1\ldots j_s}$'s. Then we can take  its cell decomposition which is almost good in the same sense as above, \ie{} obeying the conditions (i)--(iii) in one dimension lower. Proceeding as above we find \eqref{eqeq2}, and with a similar argument we find (\ref{eqeq3}).

\tocless\subsection{Shifting the $D$-contours}
\label{sec: shift}

All terms in (\ref{eqeq})-(\ref{eqeq2})-(\ref{eqeq3}) can be massaged further, by shifting the contour of integration in $D$. First consider terms like
$$
\int_{\Gamma \times S_{i_1 \dots i_p}} \mu_{Q_{i_1}, \dots, Q_{i_p}} \;.
$$
We assume that $Q_{i_1}, \dots, Q_{i_p}$ are linearly independent, otherwise the integrand just vanishes.
Recall that after (\ref{to-be-evaluated}) we chose $\delta$ such that $Q_i(\delta)\neq0$ $\forall\,i$.
If the set of indices $\{i_1, \dots, i_p\}$ contains an index $i$ such that $Q_i(\delta) < 0$, the integration domain $S_{i_1 \dots i_p}$ can be shrunk around $H_i$ keeping the integrand finite: comparing with (\ref{one-loop with D}), the real part of the denominator remains $\geq |\delta|$ without developing divergences. In this case the integral vanishes in the $\lim_{\e,\varepsilon\to0}$. Thus in all summations we can restrict to the indices $i$ such that $Q_i(\delta)>0$. In this case we do have divergences, and we cannot take the limit yet.

Then we would like to continuously deform the contour $\Gamma$ in such a way to modify the imaginary shift from $\delta$ to a new one with $Q_{i=i_1, \dots, i_p}(\im \Gamma')<0$. In general the imaginary shift can be continuously deformed (since the integrand is meromorphic in $D$), unless we hit poles. In our case $\mu_{Q_{i_1}, \dots, Q_{i_p}}$ has poles along $Q_i(D) = 0$ for $i\in\{i_1, \dots, i_p\}$. We deform the contour and along the way we pick various residues around $Q_i(D) = 0$ for $i\in \{i_1, \dots, i_p\}$:
\be
\Gamma = \Gamma_{\sli_1 \dots \sli_p} + \Gamma_{i_1 \sli_2 \dots \sli_p} + \ldots + \Gamma_{\sli_1 i_2 \dots i_p} + \Gamma_{i_1 \dots i_p}
\ee
(in homology) where we have a sum of $2^p$ terms in which each index appears either slashed or not. The contours $\Gamma_{i_1\dots i_p \slj_1 \dots \slj_q}$ have been defined in section \ref{sec: cyclesinD} and contain some arbitrariness. For the last term it will be important to choose $\delta_{i_1\dots i_p}$ such that $Q_j(\delta_{i_1\dots i_p})\neq 0$ for all $j\not\in\{i_1\dots i_p\}$, as we did for $\delta$.

If the hyperplane arrangement is \emph{projective}---as we imposed before (\ref{main formula})---the contours $\Gamma_{\dots}$ with some slashed indices make the contour in the integral $\int_{\Gamma_{\dots} \times S_{i_1 \dots i_p}} \mu_{Q_{i_1},\dots, Q_{i_p}}$ shrinkable. A possible danger is that, as $S_{i_1 \dots i_p}$ approaches another hyperplane $H_\bj$, it might happen that $Q_\bj(\im \Gamma_{\dots}) > 0$. The integrand $\mu_{Q_{i_1},\dots, Q_{i_p}}$ has poles only along $Q_i(D) =0$ for $i\in\{i_1, \dots, i_p\}$, so we can freely change the imaginary shift in the other directions of $\fh$. Such a shift does not have to be constant,%
\footnote{One may think that the very same trick can be used at the beginning, getting $Z_{T^2}=0$. This is not true. We started with $Z_{T^2} = \int_{\Gamma \times (\fM \setminus \Deps)} \mu$. At that point $\Gamma$ can be arbitrarily shifted, even as a function of $u$. However to proceed we wrote $\mu = \dd\mu_{Q_i}$ for some $i$, and in order to apply Stokes' theorem we need $\dd\mu_{Q_i}$ to be regular on $\Gamma \times (\fM \setminus \Deps)$. This forces us to choose $\im\Gamma$ such that it does not cross any hyperplane $Q_i(D) = 0$ at any point of $\fM\setminus \Deps$: in particular $\im\Gamma$ must be constant, or at least confined within a single chamber in $D$-space.}
therefore if the charges $Q_{i_1}, \dots, Q_{i_p}, Q_\bj$ involved at the intersection $S_{i_1\dots i_p\bj}$ lie on a common half of $\fh^*$, we can arrange that close to $H_\bj$ we have $Q_\bj(\im \Gamma_{\dots}) < 0$. We conclude that
\be
\label{shift 1}
\lim_{\e,\varepsilon\to0} \int_\intd{-3pt}{\Gamma \times S_{i_1 \dots i_p}} \hspace{-1cm} \mu_{Q_{i_1}, \dots, Q_{i_p}} = \bigg[ \prod_{i\in \{i_1, \dots, i_p\}} \Theta\big( Q_i(\delta) \big) \bigg] \, \lim_{\e,\varepsilon\to0} \int_\intd{-3pt}{\Gamma_{i_1 \dots i_p} \times S_{i_1 \dots i_p}} \hspace{-1cm} \mu_{Q_{i_1}, \dots, Q_{i_p}} \;.
\ee
We used the step function $\Theta(x)$, equal to $x$ if $x\geq 0$ and zero otherwise,

Next consider terms like
$$
\int_{\Gamma_{i_1\dots i_p} \times S_{i_1 \dots i_p j_1 \dots j_q}} \mu_{Q_{i_1}, \dots, Q_{i_p}, Q_{j_1}, \dots, Q_{j_q}} \;,
$$
that might come from (\ref{eqeq2})-(\ref{eqeq3}) (the previous case was $p=0$). They can be processed in a similar way as above. First we can restrict to the terms with $Q_j(\delta_{i_1 \dots i_p}) > 0$ for all $j \in \{j_1 \dots j_q\}$, otherwise the contour $S_{i_1\dots i_pj_1 \dots j_q}$ is shrinkable.
Then we can modify the contour $\Gamma_{i_1 \dots i_p}$ to a new contour such that $Q_j(\im \Gamma_{\dots}) < 0$ for $j\in\{j_1,\dots,j_q\}$, but as we do that we pick up various residue terms:
\be
\Gamma_{i_1 \dots i_p} = \Gamma_{i_1 \dots i_p \slj_1 \dots \slj_q} + \Gamma_{i_1 \dots i_p j_1 \slj_2 \dots \slj_q} + \ldots + \Gamma_{i_1 \dots i_p \slj_1 j_2 \dots j_q} + \Gamma_{i_1 \dots i_p j_1 \dots j_q}
\ee
where the sum is over $2^q$ terms. The only term that gives non-vanishing contribution in the limit is the last one:
\be
\label{shift 2}
\lim_{\e,\varepsilon\to0} \int_\intd{-3pt}{\Gamma_{i_1 \dots i_p} \times S_{i_1 \dots i_pj_1 \dots j_q}} \hspace{-2.5cm} \mu_{Q_{i_1}, \dots, Q_{i_p}, Q_{j_1}, \dots, Q_{j_q}} = \bigg[ \prod_{j\in \{j_1, \dots, j_q\}} \Theta\big( Q_j(\delta_{i_1\dots i_p}) \big) \bigg] \, \lim_{\e,\varepsilon\to0} \int_\intd{-3pt}{\Gamma_{i_1 \dots i_pj_1 \dots j_q} \times S_{i_1 \dots i_pj_1 \dots j_q}} \hspace{-3.2cm} \mu_{Q_{i_1}, \dots, Q_{i_p}, Q_{j_1}, \dots, Q_{j_q}} \;.
\ee

\tocless\subsection{The final formula}
\label{sec: iterated}

Combining the formul\ae{} \eqref{eqeq}-\eqref{eqeq3} describing the application of Stokes' theorem, with (\ref{shift 1}) and (\ref{shift 2}) from the shift of the $D$-contours, we can obtain the final formula.

Before that, however, let us discuss the choice of the vectors $\delta$ and $\delta_{i_1 \dots i_p}$ characterizing the imaginary shifts. So far we have let these choices be quite arbitrary, except for some constraints like (\ref{vectors delta i}). We will now narrow the arbitrariness. Pick a covector
\be
\eta \in \fh^* \;,
\ee
generic enough so that $\eta \not\in \Cone_\text{sing}(\sQ)$. Then we will choose the vectors $\delta$, $\delta_{i_1 \dots i_p}$ such that they satisfy the following conditions:
\begin{enumerate}
\item They are small enough that the integrand $\mu$ remains non-singular over $\Gamma \times \fM \setminus \Deps$, and the forms $\mu_{Q_{i_1}, \dots, Q_{i_p}}$ remain non-singular over $\Gamma_{i_1\dots i_p} \times S_{i_1\dots i_p}$, as the imaginary shifts in $\fh_\bC$ are turned on.
\item They satisfy the defining condition
    \be
    \label{vectors delta i bis}
    Q_{i_1}(\delta_{i_1\dots i_p}) = \cdots = Q_{i_p}(\delta_{i_1\dots i_p}) = 0 \;,
    \ee
    but are generic enough so that $Q_j(\delta_{i_1\dots i_p}) \neq 0$ for all $j \not\in \{i_1, \dots, i_p\}$, including the case $p=0$.
\item They satisfy the positivity condition
    \be
    \label{positivity condition}
    \eta(\delta) > 0 \;,\qquad\qquad\qquad \eta(\delta_{i_1 \dots i_p}) > 0 \;.
    \ee
\end{enumerate}
The newly introduced constraint 3. will be used repeatedly inside the inductive arguments  below.
We still have a huge arbitrariness in the choice of the vectors, but we will see that the final formula will only depend on $\eta$ and not on the vectors $\delta_{i_1\dots i_p}$ themselves.

Now, let us introduce the notation
\be
P_{i_1 \dots i_p} \,\equiv\, \lim_{\e,\varepsilon\to0} \int_\intd{-3pt}{\Gamma_{i_1 \dots i_p} \times S_{i_1 \dots i_p}} \hspace{-1cm} \mu_{Q_{i_1}, \dots, Q_{i_p}} \;,
\ee
including $P = \lim \int_{\Gamma \times (\fM \setminus \Deps)} \mu$ for $p=0$. Notice that $Z_{T^2} = \frac1{|W|}P$ is our goal.
Combining the formula (\ref{eqeq}) from Stokes' theorem with (\ref{shift 1}) for the shift of $D$-contours, we get the compact expression
\be
\label{formula P}
P = \sum_{m=1}^r \bigg[ -  \sum_{i_1 < \dots < i_m} \Theta\big( Q_{i_1}(\delta)\big) \cdots \Theta\big( Q_{i_m}(\delta)\big) \, P_{i_1\dots i_m} \bigg] \;.
\ee
Each of the integrals $P_{i_1\dots i_n}$ can be further massaged, combining (\ref{eqeq3}) with (\ref{shift 2}):
\be
\label{formula definition}
P_{i_1\dots i_n} = \sum_{m=n+1}^r \bigg[ -  \sum_{i_{n+1} < \dots < i_m} \Theta\big( Q_{i_{n+1}}(\delta_{i_1 \dots i_n})\big) \cdots \Theta\big( Q_{i_m}(\delta_{i_1 \dots i_n})\big) \, P_{i_1 \dots i_m} \bigg]
\ee
for $n=0, \dots, r-1$. In fact the expression in (\ref{formula P}) is just the special case $n=0$. By successive substitutions of (\ref{formula definition}) into (\ref{formula P}) one finds the final expression, which is a sum of the form
\be
\label{fobar}
P = \sum_{i_1 < \dots < i_r} c_{i_1\dots i_r} P_{i_1\dots i_r}
\ee
in terms of coefficients $c_{i_1\dots i_r}$ that we would like to determine.

It will prove convenient to prove a more general formula than (\ref{fobar}), namely:
\be
\label{formula to prove}
P_{i_1\dots i_n} = (-1)^{r-n} \sum_{i_{n+1} < \dots < i_r} \Bigg[ \prod_{j \,\in\, \{i_{n+1}, \dots, i_r\}} \Theta\big( Q_j(\delta_{i_1 \dots \widehat\jmath \dots i_r})\big) \Bigg] \, P_{i_1 \dots i_r} \;.
\ee
We will prove it by induction in $r-n$, from $n=r-1$ to $n=0$; then $n=0$ is the desired result.

The first step of the induction is at $n=r-1$:  the formula (\ref{formula to prove}) for $P_{i_1\dots i_{r-1}}$ just coincides with the expression (\ref{formula definition}) from the application of Stokes's theorem. Then we proceed by induction. Take the expression (\ref{formula definition}) of $P_{i_1\dots i_n}$, substitute (\ref{formula to prove}) for all terms $P_{i_1\dots i_m}$ with $m>n$, then subtract the expression (\ref{formula to prove}) to be proven. We get:
\begin{multline}
P_{i_1 \dots i_n} \Big|_{(\ref{formula definition})} - P_{i_1\dots i_n} \Big|_{(\ref{formula to prove})} = \\
= \sum_{m=n}^r (-1)^{r - m+1} \sum_{i_{n+1} < \dots < i_m} \Theta \big(Q_{i_{n+1}}(\delta_{i_1 \dots i_n})\big) \cdots \Theta\big(Q_{i_m}(\delta_{i_1 \dots i_n})\big) \\
\sum_{i_{m+1} < \dots < i_r} \Theta\big(Q_{i_{m+1}}(\delta_{i_1 \dots \widehat{i_{m+1}} \dots i_r})\big) \cdots \Theta\big(Q_{i_r}(\delta_{i_1 \dots i_{r-1}})\big) \, P_{i_1 \dots i_r}  \;.
\end{multline}
Notice that $-P_{i_1\dots i_n} \big|_{(\ref{formula to prove})}$ provides the term $m=n$.
Recalling that $P_{i_1 \dots i_r}$ vanishes if two indices are equal, if we expand the summations we find that each monomial has degree $r-n$ in $\Theta$, it contains all $Q_i$'s with $i \in \{i_{n+1} \dots i_r\}$, and $m-n$ (running from $0$ to $r-n$) of them have argument $\delta_{i_1\dots i_n}$ while the other $r-m$ have argument $\delta_{i_1 \dots \widehat\jmath \dots i_r}$ with $j \in \{i_{m+1}, \dots, i_r\}$. In fact the expression on the right-hand side equals
\be
\label{formula to vanish}
-  \sum_{i_{n+1} < \dots < i_r} \bigg\{ \prod_{j \,\in\, \{i_{n+1}, \dots, i_r\}} \Big[ \Theta\big(Q_j(\delta_{i_1 \dots i_n})\big) - \Theta\big(Q_j(\delta_{i_1 \dots \widehat\jmath \dots i_r})\big)  \Big] \bigg\} \, P_{i_1 \dots i_r} \;.
\ee
Consider a single summand $T_{i_1\dots i_r} = \prod_{j\in\{i_{n+1}, \dots, i_r\}} \big[ \Theta(Q_j(\delta_{i_1 \dots i_n})) - \Theta(Q_j(\delta_{i_1 \dots \widehat\jmath \dots i_r})) \big] P_{i_1 \dots i_r}$ in (\ref{formula to vanish}), for fixed $\fP = \{i_1, \dots, i_r\}$. Clearly $T_{i_1\dots i_r}$ vanishes unless the covectors $\{Q_j\}_{j \in \fP}$ are linearly independent, because of $P_{i_1\dots i_r}$. Then assume that $\{Q_j\}_{j\in\fP}$ is a basis and decompose
\be
\label{eta decomposition}
\eta = \sum\nolimits_{j\in\fP} b_\fP^j \, Q_j \;.
\ee
The subscript reminds us that the vector $b_\fP^j$ depends on $\fP$. The positivity condition (\ref{positivity condition}) together with (\ref{vectors delta i bis}) gives $\eta(\delta_{i_1\dots \widehat\jmath\dots i_r}) = b_\fP^j \, Q_j(\delta_{i_1\dots \widehat\jmath\dots i_r})>0$ for all $j\in \fP$, and
\be
\eta(\delta_{i_1\dots i_n}) = \sum_{j \,\in\, \fP \setminus\{i_1 \dots i_n\}} b_\fP^j \, Q_j(\delta_{i_1\dots i_n}) > 0
\ee
which implies that at least for one value of $j$, the corresponding term in the summation is positive. For such a $j$, we conclude that $\Theta(Q_j(\delta_{i_1 \dots i_n})) - \Theta(Q_j(\delta_{i_1 \dots \widehat\jmath \dots i_r})) = 0$. In turn, this implies that $T_{i_1\dots i_r} = 0$ and (\ref{formula to vanish}) vanishes completing the proof of (\ref{formula to prove}).

\

Now consider (\ref{formula to prove}) with $n=0$: $P = \sum_{\fP = \{i_1<\dots< i_r\}} \Theta_{\fP,\eta} \, P_{i_1\dots i_r}$ where
\be
\Theta_{\fP,\eta} = \begin{cases} \prod_{j \in \fP} \Theta\big( Q_j(\delta_{i_1 \dots \widehat\jmath \dots i_r}) \big) & \text{if } \{Q_j\}_{j\in \fP} \text{ are linearly independent} \\
0 &\text{otherwise.} \end{cases}
\ee
This is equivalent to (\ref{formula to prove}) because if $\{Q_j\}_{j\in \fP}$ are linearly dependent then $P_{i_1\dots i_r}=0$. Note that $\Theta_{\fP,\eta}$ depends on $\eta$ through the vectors $\delta_{i_1 \dots \widehat\jmath \dots i_r}$, while $P_{i_1\dots i_r}$ does not depend on it because $\delta_{i_1\dots i_r} = 0$ in the linearly independent case.
For any fixed linearly independent $\fP$, as in (\ref{eta decomposition}) let us decompose $\eta$ in the basis $\{Q_j\}_{j\in \fP}$. The positivity condition $\eta(\delta_{i_1\dots \widehat\jmath\dots i_r}) >0$ implies that $\Theta\big(Q_j(\delta_{i_1\dots \widehat\jmath\dots i_r})\big) = \Theta(b_\fP^j)$. Therefore we can rewrite $\Theta_{\fP,\eta} = \prod_{j\in\fP} \Theta(b^j_\fP)$ which is equivalent to
\be
\Theta_{\fP,\eta} = \begin{cases} 1 & \text{if } \eta \in \Cone(Q_{i_1}\dots Q_{i_r}) \\ 0 & \text{otherwise.} \end{cases}
\ee
This is precisely the factor appearing in the JK residue. Next let us analyze $P_{i_1\dots i_r}$. Assuming that $\{Q_j\}_{j\in \fP}$ are linearly independent, $\Gamma_{i_1\dots i_r} \simeq T^r$ is a middle-dimensional torus encircling $D=0$. Moreover $\mu_{Q_{i_1},\dots,Q_{i_r}}$ presents, close enough to $D=0$, only  poles along the $r$ singular hyperplanes $Q_{j\in \fP}(D) = 0$, and $\int_{\Gamma_{i_1\dots i_r}}$ computes the residue at $D=0$. We conclude that
\be
P_{i_1\dots i_r} = (-2\pi c)^r \lim_{\e,\varepsilon\to0} \int_{S_{i_1\dots i_r}} g \Big|_{D=0} \, \dd^r u = (-2\pi c)^r \lim_{\e,\varepsilon\to0} \int_{S_{i_1\dots i_r}} Z_\text{1-loop} \;.
\ee

To proceed further, let us first assume that the hyperplane arrangement $\fM_\text{sing} \subset \fM$ is \emph{non-degenerate}, in other words that at all points $u_* \in \fM_\text{sing}^*$ the number of intersecting hyperplanes is exactly $r$, and never larger. Then $S_{i_1\dots i_r}$ is a collection of  tori $T^r$, each encircling one point $u_* \in \fM_\text{sing}^*$ at the intersection of $\{H_j\}_{j\in\fP}$, and $P_{i_1\dots i_r}$ computes the residue of $Z_\text{1-loop}$ at those $u_*$. At this stage the limit $\e,\varepsilon\to0$ is trivially taken since there is no dependence any longer on them. We get
\be
P = (-2\pi c)^r \sum_{\fP = \{i_1, \dots, i_r\}} \Theta_{\fP,\eta} \int_\intd{-2pt}{S_{i_1\dots i_r}} \hspace{-.5cm} Z_\text{1-loop} = (-4\pi^2 ic)^r \sum_{u_* \in \fM_\text{sing}^*} \JKres_{u=u_*}\big( \sQ(u_*),\eta \big) \; Z_\text{1-loop} \;.
\ee
The constant $c$ can be fixed to $i/4\pi^2$ by comparing with a single rank-1 example. We thus reproduce (\ref{main formula}).

The general case of a \emph{degenerate} arrangement, in which at some point $u_* \in \fM_\text{sing}^*$ more than $r$ hyperplanes intersect, is more delicate. The main difference is that, in general, $S_{i_1\dots i_r}$ does not have the topology of $T^r$ (it is not even a closed manifold) and therefore $\int_{S_{i_1\dots i_r}}$ is not an iterated residue. What happens is that only linear combinations of the $S_{i_1\dots i_r}$'s, dictated by $\Theta_{\fP,\eta}$, are closed integration cycles. Instead of proving this directly, we notice that $\cI = \lim_{\varepsilon\to0} \sum_{\fP} \Theta_{\fP,\eta} \int_{S_{i_1 \dots i_r}}$ is a linear functional, therefore it suffices to check that it behaves as the JK residue on a basis of rational forms, in particular that (\ref{SVcondition}) holds. Suppose we apply $\cI$ to a meromorphic form that does not really have singularities along the hyperplanes $\{H_\bi\}$. Then we can take the $\varepsilon\to0$ limit by shrinking the neighborhoods $\Deps(H_\bi)$ first, and then the other ones. When we take the first limit, the integrals over all $S_{i_1\dots i_r}$'s containing some of the indices $\bi$ vanish; if the remaining hyperplanes form a non-degenerated arrangement, the remaining $S_{i_1\dots i_r}$'s becomes closed tori after the first limit. This shows that $\cI$ matches with the JK residue on the basic forms in (\ref{SVcondition}).

We conclude by noticing that despite we have used the arbitrary covector $\eta \in \fh^*$ in our manipulations, the final result (\ref{main formula}) does not depend on $\eta$ by construction. In the integral (\ref{D reinstated}) we started with, $D$ is integrated over $\fh$, there are no imaginary shifts $\delta_{i_1\dots i_p}$ nor $\eta$. In that expression we can freely shift the $D$-contour in $\fh_\bC$ without affecting the result, because there are no poles around $D=0$ at this stage. Those poles in the $D$-space are introduced by the application of Stokes' theorem.

\section{Examples}
\label{sec: examples}

To illustrate how to use the formula (\ref{main formula}), we present here various examples of increasing complexity.

\subsection{K3}

Let us begin with the example of an Abelian theory with non-degenerate singularities in $\fM$ (\ie{} more than $r$ hyperplanes never meet at one point in $\fM$ of complex dimension $r$).

Consider an $\cN{=}(2,2)$ model with gauge group $U(1)^2$, six chiral multiplets $P,X_{1,2}, Y_{1,2,3}$ with charges
$$
\begin{array}{c|cccccc|c}
& P & X_1 & X_2 & Y_1 & Y_2 & Y_3 & \text{FI} \\
\hline
U(1)_1 & -2 & 1 & 1 & 0 & 0 & 0 & \xi_1 \\
U(1)_2 & -3 & 0 & 0 & 1 & 1 & 1 & \xi_2 \\
R & 2 & 0 & 0 & 0 & 0 & 0 &
\end{array}
$$
and superpotential $W = P\, f(X,Y)$, where $f$ is a homogeneous polynomial of degree $(2,3)$ in $(X,Y)$. In the geometric phase $\xi_{1,2}>0$ the low-energy theory is a conformal non-linear sigma model on an elliptically fibered K3 defined by the curve $f(X,Y) = 0$ in $\CP^1 \times \CP^2$.

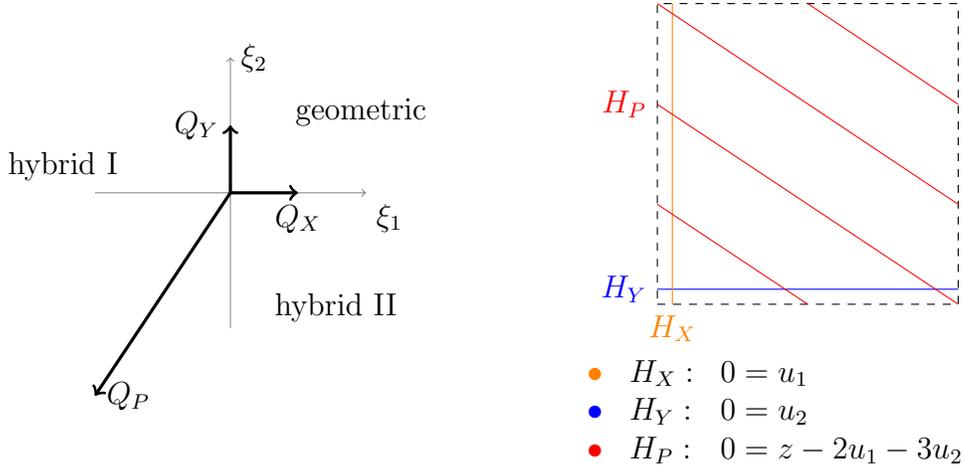
\begin{figure}[t]
\hfill
\begin{tikzpicture}[scale=.9]
\draw [->, very thin, gray] (0,-2)--(0,2) node [right, black] {$\xi_2$} ; \draw [->, very thin, gray] (-2,0) -- (2,0) node [below right, black] {$\xi_1$};
\draw [->, very thick] (0,0)--(1,0) node [below] {$Q_X$};
\draw [->, very thick] (0,0)--(0,1) node [left] {$Q_Y$};
\draw [->, very thick] (0,0)--(-2,-3) node [right] {$Q_P$};
\node [right] at (.8,1.2) {geometric};
\node [right] at (.5,-1.7) {hybrid II};
\node [left] at (-1.5,.4) {hybrid I};
\end{tikzpicture}
\hfill
\begin{minipage}[b]{.35\textwidth}
\begin{center}
\begin{tikzpicture}[scale=1]
\draw[dashed] (-.2,-.2) rectangle (3.8,3.8);
\draw[blue] (-.2,0) node[anchor=east] {$H_Y$} -- (3.8,0);
\draw[orange] (0,-.2) node[anchor=north] {$H_X$} -- (0,3.8);
\draw[red] (-.2,1.13)--(1.8,-.2);
\draw[red] (-0.2,2.46) node[anchor=east] {$H_P$} -- (3.8,-.2);
\draw[red] (-.2,3.8) -- (3.8,1.13);
\draw[red] (1.8, 3.8) -- (3.8,2.46);
\end{tikzpicture}

$
\begin{array}{cll}
\textcolor{orange}{\bullet} & H_X: & 0 = u_1 \\
\textcolor{blue}{\bullet} & H_Y: & 0 = u_2 \\
\textcolor{red}{\bullet} & H_P: & 0 = z-2u_1-3u_2
\end{array}$
\end{center}
\end{minipage}
\hfill \,
\caption{Elliptically fibered K3. Left: charge covectors in $\fh^*$, with the three phases indicated. Right: a real slice of the singular hyperplanes in $\fM$.
\label{fig: K3}}
\end{figure}

The one-loop determinant is
\be
Z_\text{1-loop}(\tau,z,u_1, u_2) = \bigg[ \frac{2\pi\eta(q)^3}{\theta_1(q,y^{-1})} \bigg]^2 \frac{\theta_1(q, x_1^{-2} x_2^{-3})}{\theta_1(q,yx_1^{-2} x_2^{-3})} \bigg[ \frac{\theta_1(q, y^{-1} x_1)}{\theta_1(q,x_1)} \bigg]^2 \bigg[ \frac{\theta_1(q, y^{-1} x_2)}{\theta_1(q, x_2)} \bigg]^3 \dd u_1 \wedge \dd u_2 \,.
\ee
The singularities in $\fM$, parametrized by $(u_1,u_2)$, are along the hyperplanes
\be
H_P = \{z - 2u_1 - 3 u_2 = 0\} \;,\qquad H_X = \{ u_1 = 0 \} \;,\qquad H_Y = \{ u_2 = 0 \} \;,
\ee
where the identifications in $\fM$ are understood.
In figure \ref{fig: K3} left we draw the charge covectors $Q_i \in \fh^*$ with the phases of the model; on the right we draw a real slice of the hyperplanes in $\fM$. In particular all intersections are non-degenerate, thus for any choice of $\eta \in \fh^*$ the cycle to use in the JK residue is simply the one of the iterated residue.

For each phase of the Abelian model, that is for each chamber in $\fh^*$, we get a different representation of the elliptic genus. In the geometric phase we get contribution from the intersection of $H_X$ and $H_Y$, \ie{} $u_1=u_2=0$. The elliptic genus is
\be
\label{K3 genus 2-param geometric}
Z_{T^2}(q,y) = \bigg[ \frac{\eta(q)^3}{i\, \theta_1(q,y^{-1})} \bigg]^2 \oint_\intd{-2pt}{u_1 = u_2 = 0} \hspace{-1.2cm} \dd u_1\, \dd u_2\, \frac{\theta_1(q, x_1^{-2} x_2^{-3})}{\theta_1(q,yx_1^{-2} x_2^{-3})} \bigg[ \frac{\theta_1(q, y^{-1} x_1)}{\theta_1(q,x_1)} \bigg]^2 \bigg[ \frac{\theta_1(q, y^{-1} x_2)}{\theta_1(q, x_2)} \bigg]^3 \;.
\ee
In the $q\to0$ limit we get
\be
Z_{T^2}(q,y) = \big( 2 y^{-1} + 20 + 2 y \big) + \cO(q) \;.
\ee
The $\chi_y$ genus of a $d$-dimensional complex manifold is $\chi_y = \sum_{p=0}^d y^{p-d/2} \sum_{q=0}^d (-1)^{p+q} h^{p,q}$, so we reproduce the $\chi_y$ genus of K3.
Similarly, we can compute the elliptic genus in the hybrid phases. For instance, consider the hybrid phase I (fig.~\ref{fig: K3}) where $H_Y \cap H_P$ contributes. Since from (\ref{SVcondition}) in this phase we have
$$
\JKres \, \frac{\dd u_1 \wedge \dd u_2}{u_2(-2u_1 -3u_2)} = \frac12 = - \frac1{(2\pi i)^2} \oint_{u_2=0}\oint_{u_1 = -3u_2/2} \, \frac{\dd u_1 \dd u_2}{u_2(-2u_1 -3u_2)} \;,
$$
we get
\begin{multline}
\label{K3 genus 2-param hybrid I}
Z_{T^2} = - \sum_{a,b=0,1} \; \frac1{(2\pi i)^2}
\oint_{u_2=0}\oint_{u_1 = \frac{z-3u_2+a+b\tau}2}
\, Z_\text{1-loop} \\
= \frac{\eta(q)^3}{2i\, \theta_1(q,y^{-1})} \sum_{a,b=0,1} y^{-b} \oint_{u_2 = 0} \hspace{-.2cm} \dd u_2 \, \bigg( \frac{\theta_1\big( \tau\big| \tfrac{-z - 3u_2 +a + b\tau}2 \big)}{\theta_1\big( \tau \big| \tfrac{z - 3u_2+a+b\tau}2 \big)} \bigg)^2 \bigg( \frac{\theta_1(\tau|-z+u_2)}{\theta_1(\tau|u_2)} \bigg)^3 \;.
\end{multline}
The other hybrid phase leads to a similar expression.

Notice that another model for K3 is a quartic hypersurface in $\CP^3$. It can be realized by an $\cN{=}(2,2)$ Abelian rank-1 theory, with chiral multiplets $(P,X_{1,2,3,4})$ with gauge charges $(-4,1)$ respectively, and a superpotential $W = P\, f(X)$ where $f$ is a homogeneous polynomial of degree 4.
If we sum over the positive poles, we obtain
\be
\label{K3 genus 1-param geom}
Z_{T^2} = \frac{\eta(q)^3}{i\, \theta_1(q,y^{-1})} \oint_{u=0} \hspace{-.2cm} \dd u \, \frac{\theta_1(q, x^{-4})}{\theta_1(q, yx^{-4})} \, \bigg( \frac{\theta_1(q,y^{-1}x)}{\theta_1(q,x)} \bigg)^4
\ee
which is the expression of the elliptic genus in the geometric phase,
and if we sum over the negative poles, we find
\be
\label{K3 genus 1-param LG}
Z_{T^2}
= \frac14 \sum_{a,b=0}^3 y^{-b} \bigg( \frac{\theta_1 \big( \tau \big| \tfrac{-3z+a+b\tau}4 \big)}{\theta_1\big( \tau \big| \tfrac{z + a + b\tau}4 \big)} \bigg)^4
\ee
which is the expression of the elliptic genus as the Landau-Ginzburg orbifold.

In fact the elliptic genus of K3 in standard form is \cite{Eguchi:1988vra}
\be
Z_{T^2}(q,y) = 8 \bigg[ \bigg( \frac{\theta_1(\tau| z + \tfrac12)}{\theta_1(\tau|\tfrac12)} \bigg)^2 + \bigg( y^{1/2} \frac{\theta_1(\tau| z + \tfrac{1+\tau}2)}{\theta_1(\tau|\tfrac{1+\tau}2)} \bigg)^2 + \bigg( y^{1/2} \frac{\theta_1(\tau| z + \tfrac\tau2)}{\theta_1(\tau|\tfrac\tau2)} \bigg)^2 \bigg] \;.
\ee
All expressions in (\ref{K3 genus 2-param geometric}), (\ref{K3 genus 2-param hybrid I}), (\ref{K3 genus 1-param geom}) and (\ref{K3 genus 1-param LG}) exactly coincide with this.

\subsection{The resolved $\bW\bP^4_{1,1,2,2,2}[8]$}

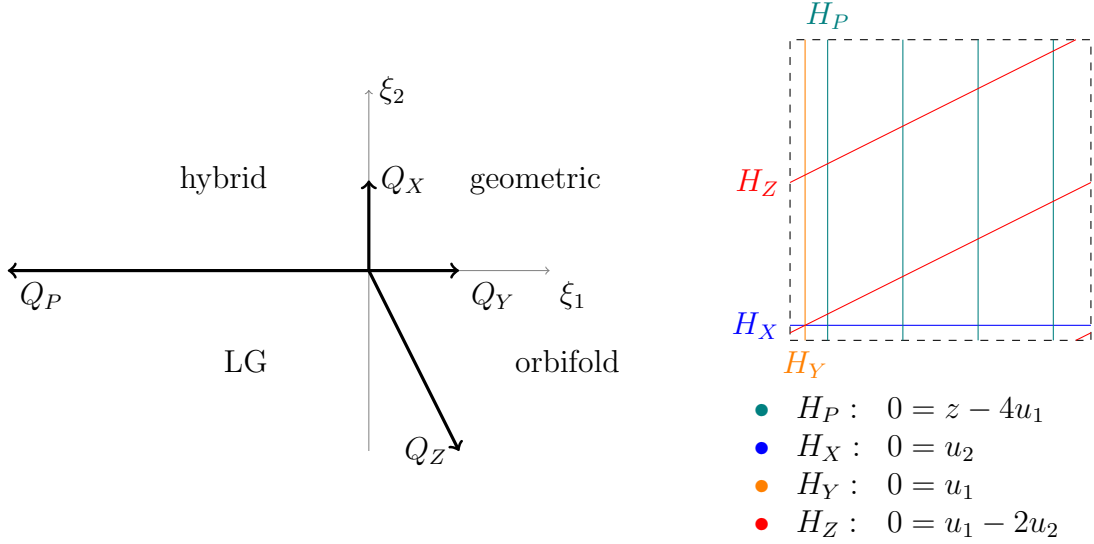
\begin{figure}[t]
\hfill
\begin{tikzpicture}[scale=1.2]
\draw [->, very thin, gray] (0,-2)--(0,2) node [right, black] {$\xi_2$} ; \draw [->, very thin, gray] (-4,0) -- (2,0) node [below right, black] {$\xi_1$};
\draw [->, very thick] (0,0)--(-4,0) node [below right] {$Q_P$};
\draw [->, very thick] (0,0)--(0,1) node [right] {$Q_X$};
\draw [->, very thick] (0,0)--(1,0) node [below right] {$Q_Y$};
\draw [->, very thick] (0,0)--(1,-2) node [left] {$Q_Z$};
\node [right] at (1,1) {geometric};
\node [right] at (1.5,-1) {orbifold};
\node [left] at (-1,-1) {LG};
\node [left] at (-1,1) {hybrid};
\end{tikzpicture}
\hfill
\begin{minipage}[b]{.35\textwidth}
\begin{center}
\begin{tikzpicture}
\draw[dashed] (-.2,-.2) rectangle (3.8,3.8);
\draw[blue] (-.2,0) node[anchor=east] { $H_X$ }-- (3.8,0);
\draw[orange] (0,-.2) node[anchor=north] { $H_Y$ }-- (0,3.8);
\draw[teal] (.3,-.2) -- (.3,3.8) node[anchor=south] {$H_P$};
\foreach \x in {1.3,2.3,3.3}
	\draw[teal] (\x,-.2) -- (\x,3.8);
\draw[red] (-.2,-0.1)--(3.8,1.9);
\draw[red] (-0.2,1.9) node[anchor=east] {$H_Z$}--(3.6,3.8);
\draw[red] (3.6,-.2)--(3.8,-.1);
\end{tikzpicture}

$\begin{array}{cll}
\textcolor{teal}{\bullet} & H_P: & 0 = z-4u_1 \\
\textcolor{blue}{\bullet} & H_X: & 0 = u_2 \\
\textcolor{orange}{\bullet} & H_Y: & 0 = u_1 \\
\textcolor{red}{\bullet} & H_Z: & 0 = u_1-2u_2
\end{array}$
\end{center}
\end{minipage}
\hfill\,
\caption{The resolved $\bW\bP^4_{1,1,2,2,2}[8]$ model. Left: charge covectors in $\fh^*$, with the four phases indicated. Right: a real slice of the singular hyperplanes in $\fM$.
\label{fig: WP model}}
\end{figure}

This is a two-parameter model analyzed in \cite{Candelas:1993dm, Morrison:1994fr}. The model has two $U(1)$ gauge fields and seven chiral multiplets $P, X_{1,2}, Y_{1,2,3}, Z$ with gauge and R-symmetry charges
\be
\label{tab: WP model}
\begin{array}{c|cccc|c}
 & P & X_{1,2} & Y_{1,2,3} & Z & \text{FI} \\
\hline
U(1)_1 & -4 & 0 & 1 & 1 & \xi_1 \\
U(1)_2 & 0 & 1 & 0& -2 & \xi_2 \\
R & 2 & 0 & 0 & 0 \\
\hline \hline
2U(1)_1 + U(1)_2 & -8 & 1 & 2 & 0 & 2\xi_1 + \xi_2
\end{array}
\ee
and a superpotential $W = P f(X,Y,Z)$ where $f$ is a weighted homogeneous polynomial.
The model has four phases as the FI parameters $\xi_{1,2}$ are varied, as shown in figure \ref{fig: WP model} left. In the geometric phase
the model describes
a hypersurface $f(X,Y,Z)=0$ in a compact toric manifold with homogeneous coordinates $X,Y,Z$, which is a smooth CY$_3$ with $h^{1,1} = 2$, $h^{2,1} = 86$, $\chi = -168$.

This CY$_3$ is the resolution of a weighted degree 8 hypersurface in a four-dimensional weighed projective space: $\bW\bP^4_{1,1,2,2,2}[8]$.  The degree 8 hypersurface in $\bW\bP^4_{1,1,2,2,2}$, birationally equivalent to the resolution, can be described by a one-parameter model we already discussed in \cite{Benini:2013nda}:%
\footnote{The resolution 2-cycle is blown down in the orbifold and Landau-Ginzburg phases of the two-parameter model: the reader can check that the expressions of the elliptic genus in those two phases mimic the geometric and Landau-Ginzburg representations, respectively, in the one-parameter model \cite{Benini:2013nda}.}
the charges are as in the last row in (\ref{tab: WP model}) and $Z$ is missing. The hypersurface has $\chi=-162$ and a genus 3 curve of $\bZ_2$ orbifold singularities which contributes the missing $\Delta\chi = -6$.

The one-loop determinant is
\be
Z_\text{1-loop} = \bigg[ \frac{2\pi\eta(q)^3}{\theta_1(q,y^{-1})} \bigg]^2 \frac{\theta_1(q,x_1^{-4})}{\theta_1(q, y x_1^{-4})} \bigg[ \frac{\theta_1(q, y^{-1} x_2)}{\theta_1(q,x_2)} \bigg]^2 \bigg[ \frac{\theta_1(q,y^{-1} x_1)}{\theta_1(q,x_1)} \bigg]^3 \frac{\theta_1(q,y^{-1} x_1 x_2^{-2})}{\theta_1(q, x_1 x_2^{-2})} \,\dd u_1 \dd u_2 \;.
\ee
The singular manifold $\fM_\text{sing}$ inside $\fM$ comprises the hyperplanes
\be
H_P = \{ z -4u_1 = 0 \} \;,\quad H_X = \{u_2 = 0\} \;,\quad H_Y = \{u_1 = 0\} \;,\quad H_Z=\{u_1 - 2u_2 = 0\}
\ee
where the identifications in $\fM$ are understood. A real slice of $\fM$ is depicted in figure \ref{fig: WP model} right. For each chamber in $\fh^*$ (figure \ref{fig: WP model} left), \ie{} for each phase of the GLSM, a choice of $\eta \in \fh^*$ in that chamber leads to a different representation of the elliptic genus. Let us consider the four representations in turn.

\paragraph{The Landau-Ginzburg phase.} For such $\eta$, the JK residue is non-vanishing at the intersection of $H_P$ and $H_Z$, which is composed of the 64 non-degenerate intersection points
$$
u_1 = \tfrac14(z+ c + d\tau) \;,\qquad u_2 = \tfrac12(u_1 + a + b \tau) \;,\qquad a,b=0,1 \;,\qquad c,d = 0, \dots, 3
$$
with a simple pole. The JK residue is $\frac1{(2\pi i)^2} \oint$. We get
\be
Z_{T^2}(q,y) = \frac18 \sum_{a,b=0}^1 \sum_{c,d=0}^3 y^{-b-d} \, \frac{\theta_1 \big( \tau \big| \frac{-7z + (4a+c) + (4b+d)\tau}8 \big)^2}{\theta_1\big( \tau \big| \frac{z + (4a+c) + (4b+d)\tau}8 \big)^2} \, \frac{\theta_1\big( \tau \big| \frac{- 3z + c + d\tau}4 \big)^3}{\theta_1 \big( \tau \big| \frac{z+c+d\tau}4 \big)^3} \;.
\ee
To compute the $\chi_y$ genus we use the $\tau\to i\infty$ (\ie{} $q \to 0$) limits
\be
\lim_{\tau \to i\infty} \frac{\theta_1 \big( \tau \big| \frac{- 7z + (4a+c) + (4b+d)\tau}8 \big)}{\theta_1\big( \tau \big| \frac{z + (4a+c) + (4b+d)\tau}8 \big)} = \begin{cases} y^{1/2} \, \dfrac{1- y^{-7/8} e^{2\pi i(4a+c)/8}}{1-y^{1/8} e^{2\pi i(4a+c)/8}} & \text{for } b,d=0 \\
y^{1/2} &\text{for } 4b+d \neq 0 \end{cases}
\ee
and
\be
\lim_{\tau \to i \infty} \frac{\theta_1\big( \tau \big| \frac{-3z + c + d\tau}4 \big)}{\theta_1 \big( \tau \big| \frac{z+c+d\tau}4 \big)} = \begin{cases} y^{1/2} \, \dfrac{1 - y^{-3/4} e^{2\pi i c/4}}{1 - y^{1/4} e^{2\pi i c/4}} & \text{for } d=0 \\ y^{1/2} & \text{for } d \neq 0 \;. \end{cases}
\ee
We get
\be
\lim_{q\to 0} Z_{T^2}(q,y) = -84\,(y^{1/2} + y^{-1/2}) \;.
\ee

\paragraph{The hybrid phase.} In this case we get contribution from the intersection of $H_P$ and $H_X$, which comprises 16 non-degenerate intersection points with non-simple poles. The JK residue is $- \frac1{(2\pi i)^2} \oint$. We get
\be
Z_{T^2} = \frac{\eta(q)^3}{4i\, \theta_1(q,y^{-1})} \sum_{a,b=0}^3 y^{-b} \bigg[ \frac{\theta_1 \big( \tau \big| \frac{-3z + a + b\tau}4 \big)}{\theta_1 \big( \tau\big| \frac{z + a + b\tau}4 \big)} \bigg]^3 \oint_{u_2=0} \hspace{-.4cm} \dd u_2\, \bigg[ \frac{\theta_1(q,y^{-1}x_2)}{\theta_1(q,x_2)} \bigg]^2 \frac{\theta_1 \big( \tau \big| \frac{-3z + a + b\tau -8u_2}4 \big)}{\theta_1 \big( \tau \big| \frac{ z + a + b\tau -8u_2}4 \big)} \;.
\ee

\paragraph{The geometric phase.} In this phase the JK residue gets contribution from $H_X \cap H_Y$ and $H_X \cap H_Z$, which is the single point $u_1 = u_2 = 0$ with a degenerate intersection of hyperplanes, and we need to apply (\ref{JK formula}). Let us spell out in some details how it works.

At $u_* = (0,0)$ the set of relevant charges is $\sQ_* = \{Q_X, Q_Y, Q_Z\}$. The set of flags with the respective vectors $\kappa_j^F$ is
\bea
\label{flags WP model}
F_1 &= \{Q_X, \bR^2\} \qquad\qquad & \kappa^{F_1} &= \{Q_X, Q_X + Q_Y + Q_Z \} \qquad\qquad & \nu(F_1) &= -1 \\
F_2 &= \{Q_Y, \bR^2\} \qquad\qquad & \kappa^{F_2} &= \{Q_Y, Q_X + Q_Y + Q_Z \} \qquad\qquad & \nu(F_2) &= -1 \\
F_3 &= \{Q_Z, \bR^2\} \qquad\qquad & \kappa^{F_3} &= \{Q_Z, Q_X + Q_Y + Q_Z \} \qquad\qquad & \nu(F_3) &= 1 \;.
\eea
For $\eta \in \Cone(Q_X,Q_Y)$, the only flag in $\cF\cL^+(\cQ_*,\eta)$ is $F_1$. Choosing a basis $\fB(F_1) = \{Q_X,Q_Y\}$ for the flag, the iterated residue is $\Res_{F_1} \omega = \Res_{u_1=0} \Res_{u_2=0} \omega_{21}$ where the latter is the $\dd u_2 \wedge \dd u_1$ component of $\omega$. We thus obtain
\be
Z_{T^2}(q,y) = \Res_{u_1=0} \; \Res_{u_2=0} \; Z_\text{1-loop}(\tau,z,u_1,u_2)
\ee
where, with a little abuse of notation, we have used $Z_\text{1-loop}$ for the $\dd u_1 \wedge \dd u_2$ component of the 2-form. Let us stress that the order in the iterated residue is crucial.

For a small number of hyperplanes, such as in this case, a faster way to get to the result is the following. Consider the 2-form
$$
\omega = \Big( \frac a{u_1 u_2} + \frac b{u_1(u_1-2u_2)} + \frac c{u_2(u_1 - 2u_2)} \Big)\, \dd u_1 \wedge \dd u_2 \;.
$$
Since $\Res_{u_1=0} \Res_{u_2=0} \omega_{12} = a+c$, this satisfies the conditions (\ref{SVcondition2}) in the geometric phase.

\paragraph{The orbifold phase.} In this phase $H_X \cap H_Z$ and $H_Y \cap H_Z$ contribute, \ie{} the four points $u_1=0$, $u_2 = (u_1 + a + b\tau)/2$ with $a,b=0,1$. The pole for $a=b=0$ sits at a degenerate intersection of three hyperplanes. The formula (\ref{JK formula}) produces two equivalent expressions depending on which sub-chamber of the orbifold phase $\eta$ sits in. If $\eta \in \Cone(Q_Y, Q_X + Q_Y + Q_Z)$ then $\cF\cL^+(\sQ_*,\eta)$ comprises the two flags $F_1$ and $F_2$ in (\ref{flags WP model}). Using $\fB(F_1)$ as before and $\fB(F_2) = \{Q_Y, Q_X\}$ we arrive at
\be
\JKres_{u_* = (0,0)}(\sQ_*,\eta) \; Z_\text{1-loop} = \Big( \Res_{u_1=0} \; \Res_{u_2=0} - \Res_{u_2 = 0} \; \Res_{u_1=0} \Big) \, Z_\text{1-loop} \;.
\ee
Alternatively, if $\eta \in \Cone(Q_Z, Q_X + Q_Y + Q_Z)$ then the only flag in $\cF\cL^+(\cQ_*,\eta)$ is $F_3$. Choosing a basis $\fB(F_3) = \{Q_Z, Q_X\}$ we get
\be
\JKres_{u_* = (0,0)}(\sQ_*,\eta) \; Z_\text{1-loop} = \Res_{u_2=0} \; \Res_{u_1= 2u_2} \; Z_\text{1-loop}
\ee
which is equivalent to the previous expression. In any case, the $q\to 0$ limit is $-81(y^\frac12 + y^{-\frac12})$.

At the other three poles for $(a,b) \neq (0,0)$ we have $\JKres = - \frac1{2\pi i}\oint$ that gives
$$
\frac{\eta(q)^3}{2i\, \theta_1(q,y^{-1})} \sum_{a,b} y^{-b} \oint_{u_1=0} \hspace{-.4cm} \dd u_1\, \frac{\theta_1(q,x_1^{-4})}{\theta_1(q, yx_1^{-4})} \, \bigg[ \frac{\theta_1\big( \tau \big| \frac{u_1 + a + b\tau - 2z}2 \big)}{\theta_1 \big( \tau \big| \frac{u_1 + a + b\tau}2 \big)} \bigg]^2 \, \bigg[ \frac{ \theta_1(q,y^{-1}x_1)}{\theta_1(q,x_1)} \bigg]^3 \;.
$$
The $q\to0$ limit of each of the three terms is $-(y^\frac12 + y^{-\frac12})$.

\

All the expressions we have obtained above coincide with the standard elliptic genus of a Calabi-Yau threefold with Euler number $\chi$ \cite{Kawai:1993jk}:
\be
\label{genus CY3}
Z_{T^2}(q,y) = \frac\chi2 \, (y^\frac12 + y^{-\frac12}) \prod_{n=1}^\infty \frac{(1-y^2q^n)(1-y^{-2}q^{-n})}{(1-yq^n)(1-y^{-1}q^{-n})} = \frac\chi2 \, \frac{\theta_1(q,y^2)}{\theta_1(q,y)} \;.
\ee

\subsection{R\o dland model}
\label{sec: ex Rodland}

The next example is a non-Abelian theory, which presents degenerate and non-projective singularities in $\fM$. Consider an $\cN{=}(2,2)$ model with $U(2)$ gauge group, seven fundamental chiral multiplets $X_i$ and a further seven chiral multiplets $P^j$ transforming in the $\det^{-1}$ representation. They are coupled through the superpotential
\be
W = \sum_{i,j,k=1}^7 A^{ij}_k P^k (X^1_i X^2_j - X^2_i X^1_j) \;,
\ee
where in $X^a_i$, $a$ is the gauge index, and $A^{ij}_k$ are generic coefficients antisymmetric in the upper indices. This model was first studied in \cite{Hori:2006dk, Hori:2011pd}, then in \cite{Jockers:2012zr}, to give a physical proof of a conjecture of R\o dland \cite{Rodland} that an incomplete intersection in $\CP^6$ and a complete intersection in the Grassmannian $Gr(2,7)$ are Calabi-Yau threefolds sitting on the same complexified K\"ahler moduli space, although they are not birationally equivalent. At large positive FI term ($\xi \gg 0$) the low-energy theory is a NLSM on the complete intersection $A^{ij}_k [X_i X_j]$ of 7 hyperplanes in the Grassmannian $Gr(2,7)$; for $\xi \ll 0$ instead one gets an \emph{incomplete} intersection in $\CP^6$ parameterized by homogeneous coordinates $P^k$, with the condition that the antisymmetric matrix $A^{ij}(P)$ has rank 4 instead of the generic rank 6. This latter variety is called the Pfaffian Calabi-Yau.
In \cite{Jockers:2012dk} the Gromov-Witten invariants of the two geometries have been extracted from the sphere partition function of \cite{Benini:2012ui, Doroud:2012xw}.

\begin{figure}[t]
$\vcenter{\hbox{
\begin{tikzpicture}[scale=1.5]
\draw [->, very thin, gray] (-1.5,0) -- (1.7,0) node [below right, black] {$u_1^*$};
\draw [->, very thin, gray] (0,-1.5)--(0,1.7) node [right, black] {$u_2^*$} ;
\draw [->, very thick, red] (0,0)--(0,1) node [right] {$Q_{X^2}$};
\draw [->, very thick, blue] (0,0)--(1,0) node [above] {$Q_{X^1}$};
\draw [->, very thick, orange] (0,0)--(-1,-1) node [above left] {$Q_P$};
\draw [->, very thick, teal] (0,0)--(1,-1) node [left] {$Q_{\sigma_+}$};
\draw [->, very thick, violet] (0,0)--(-1,1) node [left] {$Q_{\sigma_-}$};
\node [right] at (1,1) {Gr};
\end{tikzpicture}
}}$
\hfill
$\vcenter{\hbox{
\begin{tikzpicture}
\draw [dashed] (-2,-2) rectangle (2,2);
\draw [blue, thick] (0,2) -- (0,-2) node [below] {$H_{X^1}$};
\draw [red, thick] (-2,0) -- (2,0) node [right] {$H_{X^2}$};
\draw [orange, thick] (-1,2) node [above right] {$H_P$} -- (2,-1);
\draw [orange, thick] (-2,-1) -- (-1,-2) node [below left] {$H_P$};
\draw [teal, thick] (-1,-2) -- (2,1) node [right] {$H_{\sigma_+}$};
\draw [teal, thick] (-2,1) -- (-1,2) node [above left] {$H_{\sigma_+}$};
\draw [violet, thick] (-2,-1) -- (1,2) node [above] {$H_{\sigma_-}$};
\draw [violet, thick] (1,-2) node [below] {$H_{\sigma_-}$} -- (2,-1);
\end{tikzpicture}
}}$
\hfill
$\begin{array}{cll}
\textcolor{blue}{\bullet} & H_{X^1}: & u_1 = 0 \\
\textcolor{red}{\bullet} & H_{X^2}: & u_2 = 0 \\
\textcolor{orange}{\bullet} & H_P: & u_1 + u_2 = z \\
\textcolor{teal}{\bullet} & H_{\sigma_+}: & u_1 - u_2 = z \\
\textcolor{violet}{\bullet} & H_{\sigma_-} : & u_2 - u_1 = z
\end{array}
$
\caption{The R\o dland model. Left: charge covectors in $\fh^*$ (Grassmannian phase indicated). Right: a real slice of the singular hyperplanes in $\fM$.
\label{fig: Rodland}}
\end{figure}
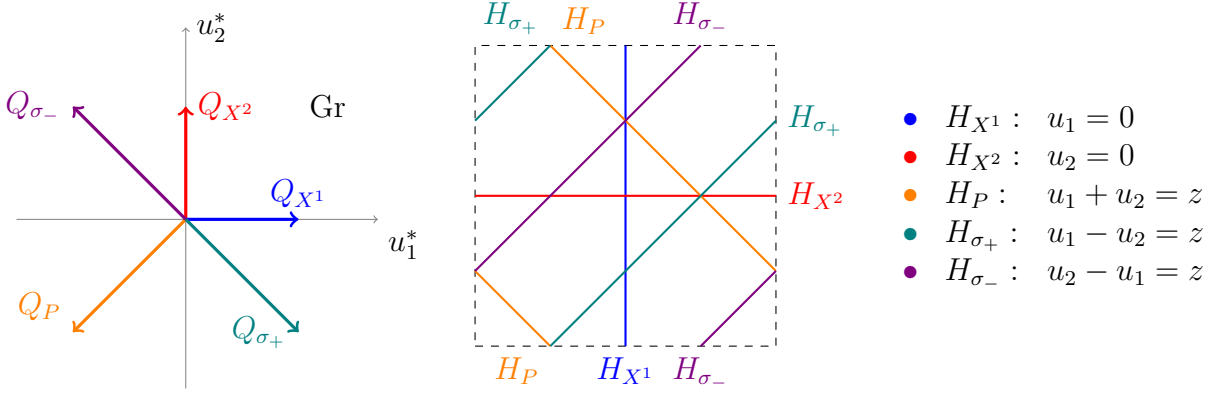

The charges are
\be
\begin{array}{c|cc|c}
& P^j & X_i & \text{FI} \\
\hline
U(2) & \rep{1}_{-2} & \square_1 & \xi \\
R & 2 & 0 &
\end{array}
\ee
The 1-loop determinant is
\begin{multline}
Z_\text{1-loop} = \frac12 \bigg[ \frac{2\pi\eta(q)^3}{\theta_1(q,y^{-1})} \bigg]^2 \frac{\theta_1(q,x_1 x_2^{-1}) \, \theta_1(q,x_1^{-1}x_2)}{\theta_1(q,y^{-1}x_1x_2^{-1}) \, \theta_1(q,y^{-1}x_1^{-1}x_2)} \\
\bigg[ \frac{\theta_1(q,y^{-1}x_1)}{\theta_1(q,x_1)} \bigg]^7 \bigg[ \frac{\theta_1(q,y^{-1}x_2)}{\theta_1(q,x_2)} \bigg]^7 \bigg[ \frac{\theta_1(q,x_1^{-1}x_2^{-1})}{\theta_1(q,yx_1^{-1}x_2^{-1})} \bigg]^7 \, \dd u_1 \wedge \dd u_2 \;.
\end{multline}
The singular hyperplanes are
\be
H_{X^1} = \{u_1 = 0\} \;,\quad H_{X^2} = \{u_2 = 0\} \;,\quad H_P = \{u_1 + u_2 = z\} \;,\quad H_{\sigma_\pm} = \{u_1 - u_2 = \pm z\} \;.
\ee
They are represented in figure \ref{fig: Rodland} together with the charge covectors in $\fh^*$.

The easiest way to perform the computation is in the Grassmannian phase, \ie{} in the chamber selected by the covector $\eta = (1,1)$. There are three intersections contributing: $H_{X^1} \cap H_{X^2}$, $H_{\sigma_-} \cap H_{X^1}$ and $H_{\sigma_+} \cap H_{X^2}$. Consider $H_{\sigma_-} \cap H_{X^1}$ first: this is a point where three linearly dependent hyperplanes meet. Unfortunately the hyperplane arrangement is not projective, and we cannot apply the JK residue directly: we need to resolve the singularity into projective ones first. We can give $P$ an R-charge $R = 2+\epsilon$ and take the limit $\epsilon \to 0$ eventually. It is easy to check that the residue at $u_1=0, u_2=z$ vanishes. A similar thing happens at $H_{\sigma_+} \cap H_{X^2}$. Hence we obtain
\be
Z_{T^2} = \frac1{(2\pi i)^2} \oint_\intd{-2pt}{u_1 = u_2 = 0} \hspace{-1cm} Z_\text{1-loop} = -49\big(y^\frac12 + y^{-\frac12} \big) + \cO(q) \;.
\ee
In fact the expression matches with (\ref{genus CY3}) with $\chi = -98$, which is the Euler number of both Calabi-Yau threefolds \cite{Batyrev}.

\subsection{Gulliksen-Neg\r{a}rd model}

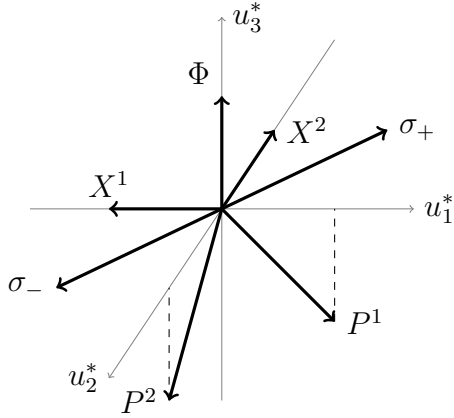
\begin{figure}[t]
\hfill
\begin{tikzpicture}[scale=1.5]
\draw [->, very thin, gray] (-1.7,0) -- (1.7,0) node [right, black] {$u_1^*$};
\draw [->, very thin, gray] (0,-1.7) -- (0,1.7) node [right, black] {$u_3^*$};
\draw [->, very thin, gray] (1, 1.5) -- (-1, -1.5) node [left, black] {$u_2^*$};
\draw [->, very thick] (0,0) -- (0,1) node [above left] {$\Phi$};
\draw [->, very thick] (0,0) -- (-1,0) node [above] {$X^1$};
\draw [->, very thick] (0,0) -- (.467,.7) node [right] {$X^2$};
\draw [->, very thick] (0,0) -- (1,-1) node [right] {$P^1$};
\draw [dashed] (1,-1) -- (1,0);
\draw [->, very thick] (0,0) -- (-.467,-1.7) node [left] {$P^2$};
\draw [dashed] (-.467,-1.7) -- (-.467,-.7);
\draw [->, very thick] (0,0) -- (1.467,.7) node [right] {$\sigma_+$};
\draw [->, very thick] (0,0) -- (-1.467,-.7) node [left] {$\sigma_-$};
\end{tikzpicture}
\hfill\,
\caption{Charge covectors of the Gulliksen-Neg\r{a}rd model in $\fh^*$. \label{fig: GN model}}
\end{figure}

This model is an $\cN{=}(2,2)$ $U(2) \times U(1)$ gauge theory with chiral multiplets $\Phi_{a=1,\dots, 8}:(\rep{1}_0,1)$, $X_{i=1,\dots,4}:(\overline\square_{-1},0)$ and $P_{i=1,\dots,4}:(\square_1,-1)$, where we have indicated the gauge charges, and superpotential
\be
W = \Tr( P^i A_{ij}^a \Phi_a X^j)
\ee
where $A^a_{ij}$ are coefficients. At low energy it flows to a NLSM on a CY$_3$ which is the locus in $\CP^7$, parametrized by the homogeneous coordinates $\Phi_a$, where the $4\times 4$ matrix $A_{ij}^a\Phi_a$ has rank $\leq 2$. The model has been studied in \cite{Jockers:2012zr}, and the Gromov-Witten invariants of the CY$_3$ have been first computed in \cite{Jockers:2012dk} from the sphere partition function of \cite{Benini:2012ui, Doroud:2012xw}. The charges under the Cartan subgroup are
\be
\begin{array}{c|ccccc|c}
& \Phi_a & X_i^1 & X_i^2 & P_i^1 & P_i^2 & \text{FI} \\
\hline
U(1)_1 & 0 & -1 & 0 & 1 & 0 & \xi_1 \\
U(1)_2 & 0 & 0 & -1 & 0 & 1 &  \xi_1 \\
U(1)_3 & 1 & 0 & 0 & -1 & -1 & \xi_2 \\
R & 0 & 0 & 0 & 2 & 2 &
\end{array}
\ee
where $U(1)_1 \times U(1)_2$ is the maximal torus of $U(2)$. In figure \ref{fig: GN model} we draw the charge covectors.

The one-loop determinant is
\begin{multline}
Z_\text{1-loop} = \frac12 \bigg[ \frac{2\pi \eta(q)^3}{\theta_1(q,y^{-1})} \bigg]^3 \frac{\theta_1(q,x_1 x_2^{-1}) \, \theta_1(q,x_1^{-1}x_2)}{\theta_1(q,y^{-1}x_1x_2^{-1}) \, \theta_1(q,y^{-1}x_1^{-1}x_2)} \, \frac{\theta_1(q,y^{-1}x_3)}{\theta_1(q,x_3)} \\
\bigg[ \frac{\theta_1(q,y^{-1}x_1^{-1})}{\theta_1(q,x_1^{-1})} \bigg]^4 \bigg[ \frac{ \theta_1(q, y^{-1}x_2^{-1})}{\theta_1(q, x_2^{-1})} \bigg]^4 \bigg[ \frac{\theta_1(q, x_1 x_3^{-1})}{\theta_1(q,yx_1 x_3^{-1})} \bigg]^4 \bigg[ \frac{\theta_1(q, x_2 x_3^{-1})}{\theta_1(q,yx_2 x_3^{-1})} \bigg]^4
\dd u_1 \dd u_2 \dd u_3\;.
\end{multline}

The easiest way to do the computation is choosing the covector $\eta = (-1,-1,1)$. Then the JK residue gets contribution only from $H_\Phi \cap H_{X^1} \cap H_{X^2}$, which is the single point $u_1 = u_2 = u_3 = 0$. We have $\JKres = \frac1{(2\pi i)^3} \oint$, therefore
\be
Z_{T^2} = \frac1{(2\pi i)^3} \oint_\intd{-3pt}{u_1 = u_2 = u_3 = 0} \hspace{-1.5cm} Z_\text{1-loop} = - 32 \big( y^\frac12 + y^{-\frac12} \big) + \cO(q) \;.
\ee
This matches with the known Euler number of the CY$_3$: $\chi = -64$ (as $h^{1,1} = 2$ and $h^{2,1} = 34$). The whole expression coincides with (\ref{genus CY3}).

\subsection{General comparison to the mathematical formula}

In the appendix A of \cite{Benini:2013nda}, we reviewed the mathematical computation of the elliptic genus of a variety $X$ when $X$ is a complete intersection in  a K\"ahler quotient $M=V/\!/G$, where $G$ is a compact group and $V$ is a representation of $G$. In this section, we show that the elliptic genus obtained  mathematically  always agrees with our path integral computation, when $G=U(1)^r$. See also \cite{Gorbounov:2003uj,MaZhou,GuoZhou,Gorbounov:2006gs}

Let us recall the mathematical computation first.
A generalized genus in the sense of Hirzebruch of an almost complex manifold $X$ is
\be
\varphi(X) = \int_X \varphi(T_\bC X) \;,
\ee
where a characteristic class $\varphi(V)$ of a vector bundle $V$ is defined in terms of its Chern roots $x_i$ by
\be
\varphi(V) = \prod_i \frac{x_i}{f(x_i)} \;.
\ee
Here $f(x)$ is a formal power series in $x$, such that $f(0)=0$ and $f'(0)$ is nonzero.
The elliptic genus is obtained by choosing
\be
f_{q,y}(x) = \frac{\theta_1\big(\tau \big| \tfrac x{2\pi i} \big)}{ \theta_1\big( \tau \big| \tfrac x{2\pi i}-z \big)} \;.
\ee
Note that in other places in this paper $x$ stands for an exponentiated chemical potential,
but in this section $x$ is a non-exponentiated chemical potential. This is to respect a standard convention in mathematics to denote the Chern roots by $x$.

We apply this definition to a variety $X$ constructed as follows. We start from a choice of $G=U(1)^r$ and its representation $V$, and consider the toric quotient $M=V/\!/G$. We denote by $\xi$  the Fayet-Iliopoulos parameter used here. Given any representation $R$ of $G$, we can construct a vector bundle $[R]$ on $M=V/\!/G$ whose fiber at a point is $R$. We take $X$ to be given by the common zeros of sections of a vector bundle $[E]$ in $M$, where $E$ is a representation of $G$.

We can use the adjunction formula $T_\bC M|_X = T_\bC X \oplus [E]|_X$ to write
\be
\varphi(X) = \int_X \frac{\varphi(T_\bC M)}{\varphi([E])} = \int_M  \varphi(T_\bC M) \, \frac{e([E])}{\varphi([E])}
\ee
where $e(\cdots)$ is the Euler class.
We note that $T_\bC M\oplus [\fg] =[V]$, where $\fg$ is a complexified Lie algebra of $G$. Therefore
\be
\label{fubar}
\varphi(X) = \int_M \frac{\varphi([V])}{\varphi([\fg])} \, \frac{e([E])}{\varphi([E])} \;.
\ee

The residue formula of Jeffrey and Kirwan \cite{JeffreyKirwan}---which originated from a conjecture by Witten \cite{Witten:1992xu}---can be stated as follows in the case of a toric quotient of a vector space \cite{BrionVergne,SzenesVergne}:
\be
\int_{V/\!/G} c([R])= \, \oint \; \prod_{i=1}^{\rank G} \; \JKres_{u=0}\big( \{v\}, \xi \big) \; \frac{\prod_{w\in R} \big( 1+w(u) \big)}{\prod_{v\in V} v(u) }
\ee
where $c(\dots)$ is the total Chern class, $u\in \fh_\bC^*$, $v$ and $w$ run over the weight vectors of the representations $V$ and $R$ respectively. Applying this formula to \eqref{fubar} we find
\be
\label{mathformula}
\varphi(X)= \, {f'(0)^{\rank G}} \; \JKres_{u=0}\big( \{v\}, \xi \big) \; \frac{ \prod_{w\in E} f\big(w(u)\big)}{ \prod_{v\in V} f\big( v(u)\big)} \; \dd u_1 \cdots \dd u_r
\ee
and $v$ and $w$ run over weights of $V$ and $E$, respectively.

Let us compare this formula to the one of the elliptic genus obtained from gauge theory.
In order to translate the geometric construction above, we introduce a vector multiplet in $G=U(1)^r$, ambient-space-producing chiral multiplets $\Phi$ of R-charge $0$ in the representation $V$,
and equation-imposing chiral multiplets $P$ of R-charge $2$ in the representation $E$.
Then the integrand in \eqref{mathformula} is the same as the integrand in the gauge theory formula \eqref{main formula}.

Usually, in order to have a smooth ambient space $V/\!/G$, there is a basis of charges such that the fields $\Phi$ have positive charges, and $\xi$ is contained in a cone generated by the charge vectors of $\Phi$. Then the fields $P$ have negative charges. Letting $\eta=\xi$, there is only one $u_*\in \fM_\text{sing}^*$ which is $u_*=0$, and our gauge theory formula \eqref{main formula} reproduces the mathematical formula \eqref{mathformula}.

\subsection{Grassmannians and dualities}
\label{sec: Grassmannians}

The Grassmannian $Gr(k,N)$ of complex $k$-planes in $\bC^N$ is realized by a $U(k)$ gauge theory with $N$ flavors transforming in the fundamental. It has $SU(N)$ flavor symmetry, where the flavors are in the anti-fundamental. This model is massive, therefore we should be careful that the R-symmetry is discrete.

The one-loop determinant is
\be
Z_\text{1-loop} = \frac1{k!} \bigg( \frac{2\pi \eta(q)^3}{\theta_1(q,y^{-1})} \bigg)^k \bigg( \prod_{i\neq j}^k \frac{\theta_1(\tau|u_i - u_j)}{\theta_1(\tau|u_i - u_j - z)} \bigg) \prod_{i=1}^k \prod_{\alpha=1}^N \frac{\theta_1(\tau| u_i - \xi_\alpha - z)}{\theta_1(\tau|u_i - \xi_\alpha)} \, \dd^ku
\ee
where we have introduced flavor holonomies $e^{2\pi i \xi_\alpha}$ for $\alpha=1,\dots,N$ and with $\sum_\alpha \xi_\alpha = 0$. We will assume the $\xi_\alpha$'s to be generic. The one-loop determinant has monodromies on $\fM$:
\be
Z_\text{1-loop}(\tau,z,u_1+a+b\tau, u_2, \dots, u_k) = y^{bN} Z_\text{1-loop}(\tau,z,u_1, \dots, u_k)
\ee
for $a,b\in \bZ$. Single-valuedness requires $y^N = 1$, \ie{} $z \in \bZ/N$.

There are two classes of singular hyperplanes:
\be
H^\text{g}_{ij} = \{u_i - u_j = z\} \;,\qquad\qquad\qquad  H^\text{f}_{i\alpha} = \{u_i = \xi_\alpha\} \;,
\ee
coming from W-bosons and fundamentals respectively. For generic values of $z$, singular points which might lead to a non-vanishing residue are at the intersection of $k$ linearly independent planes (at no point more than $k$ planes intersect); in fact at least one of the planes must be from $H^\text{f}$, otherwise either there is no intersection or the planes are linearly dependent. Also notice that all poles are simple.

If we choose, for instance, $\eta = (-1, \dots, -1)$ we do not find any contribution at all, therefore $Z_{T^2}(\tau,z,\xi) = 0$. This computation however is not valid at $z=0$ because $Z_\text{1-loop}$ is ill-defined. It is also not valid if $\gcd(k,N)>1$ and $y = e^{2\pi i j/N}$ with $j$ a multiple of $n/\gcd(k,N)$: in this case there is a non-degenerate intersection of hyperplanes coming from the W-bosons whose set of charges $\sQ(u_*)$ is not projective.%
\footnote{For instance, take $k=2$, $N=4$. For $y=-1$, the hyperplanes $u_1 - u_2 - z = 0$ and $u_2 - u_1 - z=0$ coincide because $z=\frac12 \simeq -\frac12$.}
We proceed as in section 4.3 of \cite{Benini:2013nda}: we introduce an extra chiral multiplet $P$ transforming in the $\det^{-N}$ representation to cancel the R-symmetry anomaly so that we can compute at generic $z$, but we give $P$ an R-charge 1 so that it does not affect the genus (up to a sign that we neglect) as we switch the flavor holonomies off. Therefore we compute $Z_{T^2}$ for generic values of $z$, and eventually we take a limit to the allowed values $y^N = 1$.

We choose $\eta = (1,\dots,1)$. We will show that the JK residue gets contributions only from intersections of planes purely from $H^\text{f}$. For every ordered sequence $(\bar\alpha_1, \dots, \bar\alpha_k)$ we have the intersection
\be
u_1 = \xi_{\bar\alpha_1} \;,\qquad \dots\;, \qquad u_k = \xi_{\bar\alpha_k}
\ee
of $k$ planes in $H^\text{f}$. All of these contribute to the JK residue.
If two of the $\bar\alpha_i$ are equal, then $Z_\text{1-loop}$ has a double zero in the numerator from the gauge sector and the residue vanishes; we can thus restrict to ordered sequences of unequal $\bar\alpha$'s. Given an unordered sequence, for each choice of ordering we get the same residue and such a multiplicity cancels against $k!$ in $Z_\text{1-loop}$.

There are no other contributions to $Z_{T^2}$. Consider a point at the intersection of $k$ planes, taken in part from $H^\text{f}$ and $H^\text{g}$. The JK residue picks a contribution only if $\eta$ lies inside the cone generated by the charge covectors, and this happens only if all indices $1,\dots,k$ appear either in $H^\text{f}_{i\alpha}$ or at the first position in $H^\text{g}_{ij}$. Without loss of generality suppose we picked $H^\text{g}_{\bar\imath\bar\jmath}$ and $H^\text{f}_{\bar\jmath\bar\alpha}$ for some $\bar\imath, \bar\jmath$. The zero at $u_{\bar\imath} - u_{\bar\jmath} - z=0$ in the denominator of the gauge sector is then canceled by the zero at $u_{\bar\imath} - \xi_{\bar\alpha} - z=0$ in the numerator of the flavor sector, because we also have $u_{\bar\jmath} = \xi_{\bar\alpha}$, and therefore the residue is zero.

After a further cancelation between the W-boson determinants and the fundamentals with $\alpha$ in the sequence, we get:
\be
\label{genus U(k) N}
Z_{T^2}(q,y,e^{2\pi i \xi_\alpha}) = \sum_{\cI \,\in\, C(k,N)} \; \prod_{\alpha \,\in\, \cI} \; \prod_{\beta \,\not\in\, \cI} \frac{\theta_1(\tau|\xi_\alpha - \xi_\beta - z)}{\theta_1(\tau|\xi_\alpha - \xi_\beta)} \;.
\ee
The notation is that $C(k,N)$ are combinations of $k$ elements out of the first $N$ integers, and $\cI$ is one such unordered sequence. Notice that, as it should, $Z_{T^2}$ is invariant under a common shift of all $\xi_\alpha$'s. Taking the $z\to 0$ limit we simply get the Euler number of the complex Grassmannian:
\be
Z_{T^2}(q,1) = \sum_{\bar\alpha_1 < \dots < \bar\alpha_k}^N 1 = \binom{N}{k} = \chi_{Gr(k,N)} \;.
\ee
Instead using the limit
\be
\frac{\theta_1(\tau|a)}{\theta_1(\tau|b)} \quad\xrightarrow[q\to0]{}\quad \frac{e^{i\pi a} - e^{-i\pi a}}{e^{i\pi b} - e^{-i\pi b}} \, \big( 1+\cO(q) \big) \;,
\ee
in the $q\to0$ limit we get the $\chi_y$ genus:
\be
Z_{T^2}(0, y) = \binom{N}{k}_y \;.
\ee
In fact, although not manifest in (\ref{genus U(k) N}), for $y^N = 1$ all higher terms in $q$ cancel out in $Z_{T^2}$ and we have
\be
Z_{T^2}(q,y,e^{2\pi i \xi_\alpha}) \Big|_{y^N=1} = \binom{N}{k}_y \;.
\ee
The dependence on the equivariant parameters $\xi_\alpha$ drops out because the harmonic forms
representing the cohomology classes are invariant under the isometry. We expressed the result in terms of the $q$-binomial
\be
\binom{N}{k}_y = \frac{[N]_y!}{[k]_y![N-k]_y!} = \frac{\big( y^{\frac N2} - y^{-\frac N2} \big) \big( y^{\frac{N-1}2} - y^{-\frac{N-1}2} \big) \dots \big( y^\frac{N-k+1}2 - y^{-\frac{N-k+1}2} \big)}{ \big( y^\frac12 - y^{-\frac12} \big) \big( y - y^{-1} \big) \dots \big( y^\frac k2 - y^{-\frac k2} \big)} \;,
\ee
defined through the $q$-number
\be
[n]_q = \frac{q^\frac n2 - q^{-\frac n2}}{q^\frac12 - q^{-\frac12}} = q^{-\frac{n-1}2} + q^{-\frac{n-3}2} + \ldots + q^{\frac{n-3}2} + q^{\frac{n-1}2}
\ee
and the $q$-factorial $[n]_q! = [1]_q [2]_q \dots [n]_q$.

As is well-known, there is a duality between $U(k)$ with $N$ fundamentals and $U(N-k)$ with $N$ fundamentals. Indeed, given the isomorphism between $C(k,N)$ and $C(N-k,N)$, the elliptic genus in (\ref{genus U(k) N}) can be rewritten as
\be
Z_{T^2}(q,y,e^{2\pi i \xi_\alpha}) = \sum_{\tilde\cI \,\in\, C(N-k,N)} \; \prod_{\beta \,\in\, \tilde\cI} \; \prod_{\alpha \,\not\in\, \tilde\cI} \frac{\theta_1(\tau|- \xi_\beta + \xi_\alpha - z)}{\theta_1(\tau| - \xi_\beta + \xi_\alpha)}
\ee
which is the elliptic genus of $U(N-k)$ with $N$ fundamentals, transforming in the fundamental of the flavor group $SU(N)$.

\subsubsection{Adding anti-fundamentals}

\begin{figure}[t]
\begin{center}
\begin{tikzpicture}[scale=1.5]
\draw [->, very thin, gray] (-1.5,0) -- (1.7,0) node [below right, black] {$u_1^*$};
\draw [->, very thin, gray] (0,-1.5)--(0,1.7) node [right, black] {$u_2^*$} ;
\draw [->, very thick] (0,0)--(0,1) node [right] {$Q^2$};
\draw [->, very thick] (0,0)--(1,0) node [above] {$Q^1$};
\draw [->, very thick] (0,0)--(0,-1) node [right] {$\tilde Q^2$};
\draw [->, very thick] (0,0)--(-1,0) node [above] {$\tilde Q^1$};
\draw [very thick] (0,0)--(-.4,-.4) ; \draw [very thick, dashed] (-.4,-.4)--(-.8,-.8) ; \draw [->, very thick] (-.8,-.8)--(-1.2,-1.2) node [above left] {$P_s$};
\draw [->, very thick] (0,0)--(1,-1) node [left] {$\sigma_+$};
\draw [->, very thick] (0,0)--(-1,1) node [left] {$\sigma_-$};
\draw [->, blue] (0,0)--(1,1) node [right] {$\eta$};
\end{tikzpicture}
\caption{Charge covectors of $U(k)$ with fundamentals $Q$ and anti-fundamental $\tilde Q$, for $k=2$. We included the fields $P_s$ considered in section \ref{sec: U(k) conformal}, and our choice of covector $\eta$. \label{fig: U(k) Nf Na}}
\end{center}
\end{figure}
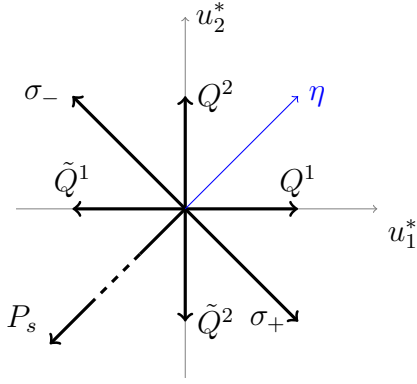

There are various generalizations that are very easy to compute. First, let us consider a theory with $N_f$ fundamentals $Q$ and $N_a$ anti-fundamentals $\tilde Q$. The theory has $SU(N_f) \times SU(N_a) \times U(1)_A$ flavor symmetry group, and the charges are
\be
\begin{array}{c|ccccc}
& U(k) & SU(N_f) & SU(N_a) & U(1)_A & U(1)_R \\
\hline
Q & \square & \overline\square & \rep{1} & 1 & 0 \\
\tilde Q & \overline\square & \rep{1} & \square & 1 & 0
\end{array}
\ee
In figure \ref{fig: U(k) Nf Na} we draw the charge covectors for the case of $U(2)$. Unless $N_f = N_a$, the R-symmetry is anomalous and we should restrict to $y^{N_a - N_f} = 1$.
We choose $\eta = (1,\dots,1)$: then the anti-fundamentals do not provide poles relevant to the JK residue. However we have to be careful about the extra chiral field $P$. If $N_f > N_a$, then $P$ does not contribute either. If $N_f = N_a$ the theory has a fixed point and we don't need $P$ at all. If $N_f < N_a$ then $P$ would contribute non-trivially: in this case we perform charge conjugation and reduce to the previous case.

So, let us assume $N_f \geq N_a$. Then we have the same poles as before, and the anti-fundamentals only contribute to the one-loop determinant. We immediately get:
\be
\label{genus U(k) Nf Na}
Z_{T^2}(q,y,e^{2\pi i \xi_\alpha}, e^{2\pi i \eta_\gamma}) = \sum_{\cI \,\in\, C(k,N_f)} \; \prod_{\alpha \,\in\, \cI} \; \prod_{\beta \,\not\in\, \cI} \; \frac{\theta_1(\tau|\xi_\alpha - \xi_\beta - z)}{\theta_1(\tau|\xi_\alpha - \xi_\beta)} \prod_{\gamma=1}^{N_a} \frac{\theta_1(\tau| - \xi_\alpha + \eta_\gamma - z)}{\theta_1(\tau | -\xi_\alpha + \eta_\gamma)}
\ee
where $\eta_\gamma$ are fugacities for $SU(N_a)$. Recall that we should impose $\sum_\alpha \xi_\alpha + \sum_\gamma \eta_\gamma = 0$ because the flavor symmetry is $S\big[U(N_f) \times U(N_a)\big]$. When evaluated at the allowed values $y^{N_a-N_f} = 1$ (for $N_f > N_a$) it simply reduces to
\be
Z_{T^2}(\tau,z,\xi_\alpha,\eta_\gamma) \Big|_{y^{N_a-N_f}=1} = y^{-kN_a/2} \binom{N}{k}_y \;.
\ee
By simple manipulations the expression in (\ref{genus U(k) Nf Na}) can be rewritten as
\begin{multline}
Z_{T^2} = \sum_{\tilde\cI \,\in\, C(N_f-k,N_f)} \; \prod_{\alpha \,\in\, \tilde\cI} \; \prod_{\beta \,\not\in\, \tilde\cI} \; \frac{\theta_1(\tau|-\xi_\alpha + \xi_\beta - z)}{\theta_1(\tau|-\xi_\alpha + \xi_\beta)} \prod_{\gamma=1}^{N_a} \frac{\theta_1(\tau| \xi_\alpha - \eta_\gamma)}{\theta_1(\tau | \xi_\alpha - \eta_\gamma + z)} \times \, \\
\,\times \prod_{i=1}^{N_a} \prod_{j =1}^{N_f} \frac{\theta_1(\tau| \eta_i - \xi_j - z)}{\theta_1(\tau|\eta_i - \xi_j)} \;.
\end{multline}
This is the elliptic genus of a theory with gauge group $U(N_f-k)$, $N_f$ fundamentals $q$, $N_a$ anti-fundamentals $\tilde q$, one extra singlet $M$ transforming in the bi-fundamental of the flavor group, and with superpotential $W = \tilde qMq$. The charges are
\be
\begin{array}{c|ccccc}
& U(N_f-k) & SU(N_f) & SU(N_a) & U(1)_A & U(1)_R \\
\hline
q & \square & \square & \rep{1} & -1 & 1 \\
\tilde q & \overline\square & \rep{1} & \overline\square & -1 & 1 \\
M & \rep{1} & \overline\square & \square & 2 & 0
\end{array}
\ee
This duality, reminiscent of four-dimensional Seiberg duality \cite{Seiberg:1994pq}, has been proposed in \cite{Benini:2012ui} and it is very similar to the three-dimensional duality discussed in \cite{Benini:2011mf}: as opposed to four dimensions, it applies to theories with different number of fundamentals and anti-fundamentals.

\subsubsection{Theories with conformal fixed points}
\label{sec: U(k) conformal}

The theories we considered before in this section are massive (unless $N_f = N_a$). We can change them into theories with a fixed point by adding fields $P_s$ in the representations $\det^{-q_s}$ with $\sum_s q_s = N_f - N_a$, and we will consider $q_s>0$ (we assume $N_f \geq N_a$). There is then no constraint on $z$. The elliptic genus of some of these theories has already been computed in \cite{Gadde:2013dda}. We assign R-charge 0 to all fields but include all flavor holonomies, so that generic R-charges are recovered by a shift of the flavor holonomies. The one-loop determinant is
\begin{multline}
Z_\text{1-loop} = \frac1{k!} \bigg( \frac{2\pi \eta(q)^3}{\theta_1(q,y^{-1})} \bigg)^k \bigg( \prod_{i\neq j}^k \frac{\theta_1(\tau|u_i - u_j)}{\theta_1(\tau|u_i - u_j - z)} \bigg) \bigg( \prod_s \frac{\theta_1(\tau| -q_s \sum u_i + \lambda_s - z)}{\theta_1(\tau| -q_s + \sum u_i + \lambda_s)} \bigg) \times\, \\
\,\times \prod_{i=1}^k \prod_{\alpha=1}^{N_f} \frac{\theta_1(\tau| u_i - \xi_\alpha + \chi - z)}{\theta_1(\tau|u_i - \xi_\alpha + \chi)} \prod_{\gamma=1}^{N_a} \frac{\theta_1(\tau| -u_i + \eta_\gamma + \chi - z)}{\theta_1(\tau| -u_i + \eta_\gamma + \chi)} \, \dd^ku \;.
\end{multline}
The charges are
\be
\begin{array}{c|cccccc}
& U(k) & SU(N_f) & SU(N_a) & U(1)_A & U(1)^s & U(1)_R \\
\hline
Q & \square & \overline\square & \rep{1} & 1 & 0 & 0 \\
\tilde Q & \overline\square & \rep{1} & \square & 1 & 0 & 0 \\
P_s & \det^{-q_s} & \rep{1} & \rep{1} & 0 & 1 & 0
\end{array}
\ee
and are represented in figure \ref{fig: U(k) Nf Na} for $k=2$. We introduced flavor holonomies $\xi_\alpha$, $\eta_\gamma$, $\chi$, $\lambda_s$ (with $\sum \xi_\alpha = \sum \eta_\gamma = 0$) for $SU(N_f) \times SU(N_a) \times U(1)_A \times U(1)^s$ respectively (this notation is slightly different than before, to make $U(1)_A$ more explicit).

We find
\begin{multline}
Z_{T^2}(\tau,z,\xi_\alpha, \eta_\gamma, \chi, \lambda_s) = \sum_{\cI \in C(k,N_f)} \bigg( \prod_{\alpha \,\in\, \cI} \prod_{\beta \,\not\in\, \cI} \; \frac{\theta_1(\tau|\xi_\alpha - \xi_\beta - z)}{\theta_1(\tau|\xi_\alpha - \xi_\beta)} \prod_{\gamma=1}^{N_a} \frac{\theta_1(\tau| \eta_\gamma - \xi_\alpha + 2\chi - z)}{\theta_1(\tau | \eta_\gamma - \xi_\alpha + 2\chi)} \bigg) \\
\,\times \prod_s \frac{\theta_1(-q_s \sum_{\alpha\in\cI} \xi_\alpha + q_s k \chi + \lambda_s - z)}{\theta_1(-q_s \sum_{\alpha\in\cI} \xi_\alpha + q_s k \chi + \lambda_s)} \;.
\end{multline}
Proceeding as before, $Z_{T^2}$ can be rewritten as
\begin{multline}
Z_{T^2} = \sum_{\tilde\cI \,\in\, C(N_f-k,N_f)} \; \bigg( \prod_{\alpha \,\in\, \tilde\cI} \; \prod_{\beta \,\not\in\, \tilde\cI} \; \frac{\theta_1(\tau|-\xi_\alpha + \xi_\beta - z)}{\theta_1(\tau|-\xi_\alpha + \xi_\beta)} \prod_{\gamma=1}^{N_a} \frac{\theta_1(\tau| \xi_\alpha - \eta_\gamma - 2\chi)}{\theta_1(\tau | \xi_\alpha - \eta_\gamma -2\chi + z)} \bigg) \times \, \\
\,\times \bigg( \prod_{i=1}^{N_a} \prod_{j =1}^{N_f} \frac{\theta_1(\tau| \eta_i - \xi_j +2\chi - z)}{\theta_1(\tau|\eta_i - \xi_j + 2\chi)} \bigg) \prod_s \frac{\theta_1( q_s \sum_{\alpha\in\tilde\cI} \xi_\alpha + q_sk\chi + \lambda_s - z)}{\theta_1(q_s \sum_{\alpha\in\tilde\cI} \xi_\alpha + q_s k \chi + \lambda_s)} \;.
\end{multline}
This is the elliptic genus of a theory with gauge group $U(N_f-k)$, $N_f$ fundamentals $q$, $N_a$ antifundamentals $\tilde q$, a singlet $M$ in the bifundamental of the flavor group, superpotential $W= \tilde q M q$, and fields $p_s$ transforming as $\det^{-q_s}$:
\be
\begin{array}{c|cccccc}
& U(N_f - k) & SU(N_f) & SU(N_a) & U(1)_A & U(1)^s & U(1)_R \\
\hline
q & \square & \square & \rep{1} & -1 & 0 & 1 \\
\tilde q & \overline\square & \rep{1} & \overline\square & -1 & 0 & 1 \\
M & \rep{1} & \overline\square & \square & 2 & 0 & 0 \\
p_s & \det^{-q_s} & \rep{1} & \rep{1} & q_sk & 1 & 0
\end{array}
\ee

\subsubsection{Adding one adjoint}

We can also consider a theory with gauge group $U(k)$, $N_f$ fundamentals, $N_a$ anti-fundamentals and one adjoint. The one-loop determinant is
\begin{multline}
Z_\text{1-loop} = \frac1{k!} \bigg( \frac{2\pi \eta(q)^3}{\theta_1(q,y^{-1})} \bigg)^k \bigg( \prod_{i\neq j}^k \frac{\theta_1(\tau|u_i - u_j)}{\theta_1(\tau|u_i - u_j - z)} \bigg) \bigg( \prod_{i,j=1}^k \frac{\theta_1(\tau|u_i - u_j + \lambda - z)}{\theta_1(\tau|u_i - u_j + \lambda)} \bigg) \times \, \\
\,\times \prod_{i=1}^k \prod_{\alpha=1}^{N_f} \frac{\theta_1(\tau| u_i - \xi_\alpha + \chi - z)}{\theta_1(\tau|u_i - \xi_\alpha + \chi)} \prod_{\gamma=1}^{N_a} \frac{\theta_1(\tau| -u_i + \eta_\gamma + \chi - z)}{\theta_1(\tau| -u_i + \eta_\gamma + \chi)} \, \dd^ku \;.
\end{multline}
We have introduced flavor holonomies $\xi_\alpha$, $\eta_\gamma$, $\chi$, $\lambda$ (with $\sum \xi_\alpha = \sum \eta_\gamma = 0$) for $SU(N_f) \times SU(N_a) \times U(1)_A \times U(1)_\Phi$.

For simplicity, let us consider first the case of $\cN{=}(4,4)$ or $\cN{=}(2,2)^*$. In this case $N_f = N_a \equiv N$ and there is a superpotential term $W = Q \Phi \tilde Q$ which imposes the constraints
\be
\label{(4,4) constraints}
2\chi + \lambda = z \;,\qquad\qquad\qquad \xi_\alpha = \eta_\alpha \quad \forall\, \alpha
\ee
from the breaking of the flavor group to $SU(N) \times U(1)_A$.
In this case it is easy to see that the only poles contributing are the same ones as before. Suppose we want to use a pole at $u_i - u_j + \lambda = 0$ from the denominator of the adjoint: such a pole cancels with a zero at $u_i - \xi_\alpha - \chi + z = 0$ from the numerator of the anti-fundamentals, using $u_j = \xi_\alpha - \chi$. We thus get, after various cancelations:
\be
\label{genus (4,4)}
Z_{T^2}(\tau,z,\lambda,\xi_\alpha) = \sum_{\cI \,\in\, C(k,N)} \; \prod_{\alpha \,\in\, \cI} \; \prod_{\beta \,\not\in\, \cI} \; \frac{\theta_1(\tau|\xi_\alpha - \xi_\beta - z)}{\theta_1(\tau|\xi_\alpha - \xi_\beta)} \; \frac{\theta_1(\tau| \xi_\alpha - \xi_\beta + \lambda)}{\theta_1(\tau|\xi_\alpha - \xi_\beta + \lambda - z)} \;.
\ee
This is the $\cN{=}(4,4)$ version of (\ref{genus U(k) N}). At $q\to 0$, in the $\lambda \to \infty$ limit we recover $U(k)$ with $N$ fundamentals.

By rewriting the sum over $\cI \in C(k,N)$ as a sum over $\tilde\cI \in C(N-k,N)$, we can rewrite the elliptic genus for $\cN{=}(4,4)$ $U(k)$ with $N$ hypermultiplets in (\ref{genus (4,4)}) as the genus of $U(N-k)$ with $N$ hypermultiplets: the precise map of parameters is
\be
Z^{\cN=(4,4)}_{U(k),\,N}(\tau,z,\xi_\alpha,\lambda) = Z^{\cN=(4,4)}_{U(N-k),\,N}(\tau,z,-\xi_\alpha,\lambda) \;.
\ee
In the geometric phase both theories flow to the NLSM on $T^*Gr(k,N)$, the cotangent bundle to the Grassmannian.

\

Finally, let us relax the superpotential and the constraints (\ref{(4,4) constraints}), but still keeping $N_f = N_a \equiv N$. Now we no longer have cancelations between the adjoint and the anti-fundamentals, therefore more poles contribute. Choosing $\eta = (1,\dots,1)$ as before, we get contributions to the JK residue from intersections of the hyperplanes
\be
H^\Phi_{ij} = \{u_i -u_j+\lambda = 0\} \qquad\qquad\qquad H^\text{f}_{i\alpha} = \{u_i -\xi_\alpha + \chi=0\} \;.
\ee
More precisely, we have to pick collections of hyperplanes such that all indices $i=\{1,\dots,k\}$ appear either in $H^\text{f}_{i\alpha}$ or at the first position in $H^\Phi_{ij}$. We can think of one such collection as defining a (possibly disconnected) graph: each $H^\text{f}_{i\alpha}$ is the root of a component, and each $H^\Phi_{ij}$ adds a segment to an existing component. If the graph has cycles then the charge covectors are linearly dependent; if a component branches in two because we used $H^\Phi_{\bi\bj}$ and \raisebox{0pt}[0pt][0pt]{$H^\Phi_{\overline{k\jmath}}$} for some $\bi,\bj,\bar k$, then we also get a zero from the numerator of the gauge sector, $u_\bi - u_{\bar k} =0$. The only contributing graphs are then disconnected chains.
Hence, the set of poles is parametrized by ordered sequences $\vec n = (n_1,\dots,n_N)$ with $n_\alpha \geq 0$ and $\sum_\alpha n_\alpha = k$. For every such sequence we have
\be
\{u_i\} = \left\{ \begin{aligned} & \xi_1 \,,\; \xi_1 - \chi , \dots, \xi_1 - (n_1-1)\chi \,, \\
&\vdots \\
& \xi_N \,,\; \xi_N - \chi , \dots, \xi_N - (n_N-1)\chi \end{aligned} \right.
\ee
where each row exists only for $n_\alpha > 0$. Taking into account the $k!$ permutations of $\{u_i\}$ leading to the same residue, we cancel the Weyl group dimension $|W|$. More compactly we can replace $\prod_{i=1}^k \to \prod_{\alpha=1}^N \prod_{m_\alpha=0}^{n_\alpha-1}$ and $u_i = \xi_\alpha - \chi - m_\alpha\lambda$. After many cancelations%
\footnote{The second fraction in (\ref{genus U(k) adj}) straightforwardly comes from the anti-fundamentals. Between W-bosons, the adjoint and the fundamentals there are many cancelations. Let us parametrize $i$ by $(\alpha,m_\alpha)$. Given all terms $(\alpha,m_\alpha),(\beta,l_\beta)$ from W-bosons---with $(\alpha,m_\alpha)\neq (\beta,l_\beta)$---those with $l_\beta \geq 1$ cancel against terms $(\alpha,m_\alpha),(\beta,l_\beta-1)$ from $\Phi$, while those with $l_\beta = 0$ cancel against terms from the fundamentals. One is left with the first fraction in (\ref{genus U(k) adj}).}
we get
\be
\label{genus U(k) adj}
Z_{T^2} = \sum_{\substack{\vec n \text{ s.t.} \\[3pt] |\vec n| = k}} \; \prod_{\alpha,\beta=1}^N \; \prod_{m_\alpha=0}^{n_\alpha-1} \; \frac{\theta_1 \big(\tau \big| \xi_\alpha - \xi_\beta + (n_\beta - m_\alpha)\lambda - z \big)} {\theta_1 \big( \tau \big| \xi_\alpha - \xi_\beta + (n_\beta - m_\alpha)\lambda \big)} \; \frac{\theta_1( \tau | -\xi_\alpha + \eta_\beta + m_\alpha\lambda + 2\chi - z)}{\theta_1( \tau | -\xi_\alpha + \eta_\beta + m_\alpha\lambda + 2\chi)}
\ee
where $|\vec n| \equiv \sum_\alpha n_\alpha$.
This expression has also been found in \cite{Gadde:2013dda}. As a check, if we set the $\cN=(4,4)$ constraints (\ref{(4,4) constraints}) then only sequences $\vec n$ with $n_\alpha \in\{0,1\}$ contribute to the sum and we reproduce (\ref{genus (4,4)}).

\subsection{$SU(k)$ with $N$ fundamentals}

Let us use our analysis of $U(k)$ theories in the previous section to obtain the elliptic genus of an $SU(k)$ theory with $N$ fundamentals.

We start from a $U(k)$ theory with $N$ fundamental chiral multiplets (in the anti-fundamental of the $SU(N)$ flavor group), together with $N$ chiral multiplets in the representation $\det^{-1}$.
Let us call this theory $U$.
We denote the flavor holonomies of the $SU(N)$ flavor symmetry by $\xi_\alpha$, $\alpha=1,\ldots,N$ with $\sum_\alpha \xi_\alpha = 0$, and of the $N$ chirals in $\det^{-1}$  by $\lambda_s$, $s=1,\ldots,N$.
The elliptic genus $Z_{U}(\tau,z,\xi_\alpha,\lambda_s)$ of this model $U$ has already been obtained in section \ref{sec: U(k) conformal}. In this case it is given by
\be
Z_{U}(\tau,z,\xi_\alpha,\lambda_s)
= \sum_{\cI \,\in\, C(k,N)} \bigg( \prod_{\alpha\,\in\,\cI} \; \prod_{\beta\,\not\in\,\cI} \frac{\theta_1(\tau | \xi_\alpha-\xi_\beta - z)}{\theta_1(\tau | \xi_\alpha-\xi_\beta)} \bigg) \prod_{s=1}^N \frac{\theta_1(\tau | -\sum_{\alpha\in\cI} \xi_\alpha + \lambda_s - z)}{\theta_1( \tau | -\sum_{\alpha\in\cI} \xi_\alpha + \lambda_s)} \;.\label{zu}
\end{equation}

Consider instead the $SU(k)$ gauge theory with $N$ fundamentals, with flavor holonomies $(\xi_\alpha,u)$ for the flavor symmetry $SU(N) \times U(1)$. The $U(1)$ part is normalized such that the baryons have charge $1$, \ie{} the fundamentals have holonomies $-\xi_\alpha + u/k$. Let us call this model $S$, and our objective is to compute its elliptic genus $Z_S(\tau,z,\xi_\alpha,u)$.

In the model $U$, when the coupling of the $SU(k)$ part is far stronger than the $U(1)$ part,
the $SU(k)$ part becomes non-perturbative first. Then, we can describe the theory as a $U(1)$ gauge theory, coupled to the $U(1)$ flavor symmetry of the  theory $S$, together with $N$ additional charge $-1$ fields $P_s$ with flavor holonomies $\lambda_s$. In this description, the elliptic genus of the model $U$ is given by
\be
Z_U(\tau,z,\xi_\alpha,\lambda_s) = \sum_{u_*} \; \JKres_{u_*} \; \frac{2\pi \eta(q)^3}{\theta_1(q,y^{-1})} \; Z_S(\tau,z,\xi_\alpha,u) \; \prod_{s=1}^N \frac{\theta_1(\tau | -u+\lambda_s - z)}{\theta_1(\tau | -u+\lambda_s)} \, \dd u \;.
\ee
As $Z_S$ has only positive poles, if we compute the residue by summing over negative poles we only pick the poles from the $P$'s (and recall a minus sign from $\JKres$):
\be
Z_U(\tau,z,\xi_\alpha,\lambda_s) = \sum_{s=1}^N Z_S(\tau,z,\xi_\alpha,\lambda_s) \prod_{a\, (\neq s)}^N \frac{\theta_1( \tau | -\lambda_s + \lambda_a - z)}{\theta_1( \tau | -\lambda_s + \lambda_a)} \;.
\ee
To extract the function $Z_S(\tau,z,\xi_\alpha,u)$, we just take a specific set of holonomies:
\be
\lambda_s = u - (s-1)\,z \;.
\ee
After a small computation we have
\be
Z_U \big(\tau,z,\xi_\alpha,\, \lambda_s = u - (s-1)z \big) =  Z_S(\tau,z,\xi_\alpha,u) \, \frac{\theta_1(\tau | Nz)}{\theta_1(\tau| z)} \;.
\ee
Plugging in \eqref{zu},  we find the desired expression:
\begin{multline}
\label{genus SU(k)}
Z_{SU(k),\, N}(\tau,z,\xi_\alpha,u) = \\
 \frac{\theta_1(\tau|z)}{\theta_1(\tau|Nz)} \; \sum_{\cI \,\in\, C(k,N)} \bigg( \prod_{\alpha\,\in\,\cI} \; \prod_{\beta\,\not\in\,\cI} \frac{\theta_1(\tau | \xi_\alpha-\xi_\beta - z)}{\theta_1( \tau | \xi_\alpha-\xi_\beta)} \bigg) \; \frac{\theta_1\big( \tau \big| -\sum_{\alpha\in\cI} \xi_\alpha + u - Nz \big)}{\theta_1 \big( \tau \big| -\sum_{\alpha\in\cI} \xi_\alpha + u \big)} \;.
\end{multline}
In the limit $z\to 0$, this yields
\be
Z_{SU(k),\,N}(\tau,z,\xi_\alpha,u) \;\stackrel{z\to 0}{\longrightarrow}\; \frac{1}{N}{N\choose k} \;.
\ee
This agrees with the result in \cite{Hori:2006dk}, for choices of $N$ and $k$ such that there is no non-compact Coulomb branch.

The expression in (\ref{genus SU(k)}) can be easily rewritten as a sum over $\tilde\cI \in C(N-k,N)$, as we did for $U(k)$ theories in section \ref{sec: Grassmannians}. We thus find equality of the elliptic genus of $SU(k)$ with $N$ fundamentals and $SU(N-k)$ with $N$ fundamentals, confirming the duality proposed in \cite{Hori:2006dk}. The precise map of parameters is
\be
Z_{SU(k),\, N}(\tau,z,\xi_\alpha,u) = Z_{SU(N-k),\, N}(\tau,z,-\xi_\alpha,u) \;,
\ee
and we remark that on both sides the baryons have charge 1.
Equality of the $S^2$-partition function for the two theories, with the same map of parameters, was shown in \cite{Benini:2012ui}.

As a simple check, the elliptic genus of $SU(k)$ with $N=k+1$ can be further rewritten as
\be
Z_{SU(k),\, k+1}(\tau,z,\xi_\alpha,u) = \prod_{\alpha=1}^{k+1} \frac{\theta_1(\tau|\xi_\alpha + u - z)}{\theta_1(\tau|\xi_\alpha+u)}
\ee
using identities of theta functions. This is easier to check at the level of $\chi_y$ genus:
\be
Z_{SU(k),\, k+1}(\tau,z,\xi_\alpha,u) \;\stackrel{q\to 0}{\longrightarrow}\;
y^{-\frac{k+1}2}\prod_{\alpha=1}^{k+1} \frac{y-e^{2\pi i(\xi_\alpha + u)}}{1-e^{2\pi i(\xi_\alpha + u)}} \;,
\ee
The ones above are the elliptic and $\chi_y$ genus of a chiral multiplet transforming in the fundamental of $SU(N)$, and with baryon $U(1)$ charge 1. This agrees with the result in \cite{Hori:2006dk} that $SU(k)$ gauge theory with $N=k+1$ fundamentals becomes a theory of $N$ free baryons in the infrared.

We notice that the trick we employed to extract the genus of the ``ungauged'' theory---here $SU(k)$---from the genus of the ``gauged'' one---here $U(k)$---only works if the matter of the ungauged theory provides only positive poles. This is consistent with the fact that a similar duality does not hold for $SU(k)$ with both fundamentals and anti-fundamentals, at least in this simple form.

\section*{Acknowledgements}
RE and YT thank the Aspen Center for Physics  for hospitality while the manuscript was finalized and were
partially supported by the NSF Grant \#1066293 during their visit there.
FB's work is supported in part by DOE grant DE-FG02-92ER-40697.
KH's work is supported in part by JSPS Grant-in-Aid for Scientific Research No. 21340109.
YT's work is supported in part by JSPS Grant-in-Aid for Scientific Research No. 25870159.
RE, KH, YT are also supported in part by WPI Initiative, MEXT, Japan at IPMU, the University of Tokyo.

\appendix

\section{Eta and theta functions}
\label{app: theta}

The Dedekind eta function is
\be
\eta(\tau)=q^{1/24}\prod_{n = 1}^\infty (1-q^n)
\ee
where $q = e^{2\pi i \tau}$ and $\im\tau > 0$. We will also write $\eta(q)$.  Its modular properties are
\be
\eta(\tau + 1) = e^{i \pi/12} \, \eta(\tau) \;,\qquad\qquad \eta \Big( - \frac1\tau \Big) = \sqrt{-i \tau} \, \eta(\tau)
\ee
and $\eta(\tau)^{24}$ is a modular form of weight 12.
The Jacobi theta function we use is
\bea
\theta_1(\tau | z) &= -i q^{1/8} y^{1/2} \prod_{k=1}^\infty (1-q^k) (1-y q^k) (1-y^{-1}q^{k-1}) \\
&= -i \sum_{n\in \bZ} (-1)^n e^{2\pi i z \left( n + \frac12 \right)} e^{\pi i \tau \left( n + \frac12 \right)^2}
\eea
where $q$ is as before and $y = e^{2\pi i z}$. We will also use the notation $\theta_1(q,y)$.

Under shifts of $z$ the Jacobi theta function transforms as
\be
\theta_1(\tau | z+a + b\tau) = (-1)^{a+b} \, e^{- 2\pi i b z - i \pi b^2 \tau} \, \theta_1(\tau|z)\label{zot}
\ee
for $a,b \in \bZ$.
Moreover
\be
\theta_1 (\tau| - z) = - \theta_1(\tau|z) \;.
\ee
The function $\theta_1(\tau|z)$ has simple zeros in $z$ at $z = \bZ + \tau \bZ$ and no poles. To compute residues it is useful to note that
\be
\theta_1'(\tau|0) = 2\pi \, \eta(q)^3
\ee
where the derivative is taken with respect to $z$. Combined with \eqref{zot} it gives the residue:
\be
\label{boo}
\frac1{2\pi i} \oint_{u \,=\, a + b\tau} \hspace{-.8cm} \dd u \quad \frac1{\theta_1(\tau|u)} = \frac{(-1)^{a+b} e^{i\pi b^2\tau}}{2\pi \eta(q)^3} \;.
\ee
The modular properties are:
\be
\theta_1(\tau+1| z) = e^{\pi i /4} \, \theta(\tau|z) \;,\qquad\qquad \theta_1 \Big( - \frac1\tau \Big| \frac z\tau \Big) = -i \, \sqrt{-i \tau} \, e^{\pi i z^2/\tau} \, \theta_1(\tau| z) \;.
\ee

\section{Supersymmetry and actions}
\label{app: susy}

Our conventions for the supersymmetry variations and the actions in Euclidean signature and the same as in \cite{Benini:2012ui, Benini:2013nda}. The multiplication for anticommuting Dirac spinors is
\be
\epsilon \psi = \psi \epsilon \equiv \epsilon^\sT C \psi = \epsilon^\alpha C_{\alpha\beta} \psi^\beta
\ee
where $C$ is the charge conjugation matrix. We take $C = \gamma_2$ (the Pauli matrix) so that $C^2 = 1$ and $C^\sT = -C$, in particular $\epsilon \gamma^\mu \psi = - \psi \gamma^\mu \epsilon$.
The chirality matrix is $\gamma_3 = -i \gamma_1\gamma_2$. In components
\be
\epsilon\psi = \epsilon^+ \psi_+ + \epsilon^- \psi_- = -i \epsilon^+ \psi^- + i \epsilon^- \psi^+ \;,
\ee
hence we see how to raise and lower indices. Finally the Fierz identity for anticommuting fermions is
\be
(\bar\epsilon \lambda_1) \lambda_2 = - \tfrac12 \big[ \lambda_1 (\bar\epsilon \lambda_2) + \gamma_3 \lambda_1 (\bar\epsilon \gamma_3 \lambda_2) + \gamma_\mu \lambda_1 (\bar\epsilon \gamma^\mu \lambda_2) \big] \;.
\ee
To go to Euclidean signature we set $x^0 = i x^2$, therefore $F_{01} = i F_{12}$. Since the flux pairs up holomorphically with the D-term in Lorentzian signature $D_L$, we define $F_{01} + i D_L = i (F_{12} + i D)$ hence $D_L = i D$.

With $\cN{=}(2,2)$ supersymmetry, first we have a vector multiplet $V_{(2,2)} = (A_\mu, \lambda, \bar\lambda, \sigma, \bar\sigma, D)$ with variations:
\bea
\delta A_\mu &= - \frac i2 \big( \bar\epsilon \gamma_\mu \lambda + \bar\lambda \gamma_\mu \epsilon \big) \\
\delta \sigma &= \bar\epsilon P_- \lambda + \bar\lambda P_- \epsilon \\
\delta\bar\sigma &= \bar\epsilon P_+ \lambda + \bar\lambda P_+ \epsilon \\
\delta\lambda &= + i \gamma_3 \epsilon \, F_{12} - \epsilon \, D - i P_- \epsilon\, [\sigma, \bar\sigma] + i \gamma^\mu P_+ \epsilon\, D_\mu \sigma + i \gamma^\mu P_- \epsilon\, D_\mu \bar\sigma \\
\delta\bar\lambda &= - i \gamma_3 \bar\epsilon\, F_{12} - \bar\epsilon\, D - i P_- \bar\epsilon\, [\sigma, \bar\sigma] + i \gamma^\mu P_- \bar\epsilon\, D_\mu\sigma + i \gamma^\mu P_+ \bar\epsilon\, D_\mu \bar\sigma \\
\delta D &= - \frac i2 \bar\epsilon \gamma^\mu D_\mu \lambda + \frac i2 D_\mu \bar\lambda \gamma^\mu \epsilon + i [\bar\epsilon P_+ \lambda, \sigma] - i [\bar\lambda P_- \epsilon, \bar\sigma] \;, \\
\eea
where
\be
P_\pm = \frac{1 \pm \gamma_3}2 \;.\label{eq:abcde}
\ee
With respect to the standard conventions, for instance of \cite{Witten:1993yc}, we shifted $D \to D + \frac i2[\sigma, \bar\sigma]$ so that the $\cN=(2,2)$ vector multiplet decomposes into $\cN=(0,2)$ multiplets more nicely.

Second we have a chiral multiplet $\Phi_{(2,2)} = (\phi, \bar\phi, \psi, \bar\psi, F, \bar F)$ with variations:
\bea
\delta \phi &= \bar\epsilon \psi \qquad\qquad &
\delta \psi &= i \gamma^\mu \epsilon\, D_\mu\phi + iP_+\epsilon\, \sigma\phi + i P_-\epsilon\, \bar\sigma \phi + \bar\epsilon \, F \\
\delta \bar\phi &= \bar\psi \epsilon &
\delta \bar\psi &= i \gamma^\mu \bar\epsilon\, D_\mu \bar\phi + i P_- \bar\epsilon\, \bar\phi \sigma + i P_+ \bar\epsilon\, \bar\phi \bar\sigma + \epsilon\, \bar F \\
&& \delta F &= \epsilon \big( i \gamma^\mu D_\mu \psi - i P_- \sigma \psi - i P_+ \bar\sigma \psi -i \lambda\phi \big) \\
&& \delta \bar F &= \bar\epsilon \big( i \gamma^\mu D_\mu \bar\psi - i P_+ \bar\psi \sigma - i P_- \bar\psi \bar\sigma - i \bar\phi \bar\lambda \big) \;.
\eea
The Yang-Mills Lagrangian is
\be
\cL_\text{YM} = \Tr \Big[ F_{12}^2 + D^2 + D_\mu \bar\sigma D^\mu \sigma + i D [\sigma, \bar\sigma] - i \bar\lambda \gamma^\mu D_\mu \lambda - i \bar\lambda P_+ [\sigma, \lambda] - i \bar\lambda P_- [\bar\sigma, \lambda] \Big] \;,
\ee
while the kinetic Lagrangian for the chiral multiplet is
\be
\cL_\text{mat} = D_\mu \bar\phi D^\mu \phi + \bar\phi \big( \bar\sigma \sigma + i D \big) \phi + \bar F F - i \bar\psi \gamma^\mu D_\mu \psi + i \bar\psi \big( P_- \sigma + P_+ \bar\sigma \big) \psi + i \bar\psi \lambda \phi + i \bar\phi \bar\lambda \psi \;.
\ee

To reduce to $\cN{=}(0,2)$ supersymmetry, we can take chiral parameters \hbox{$P_- \epsilon = P_- \bar\epsilon = 0$}. We define complex coordinates $w = x^1 + i x^2$, $\bar w = x^1 - ix^2$, so that $\gamma^w \epsilon = \gamma^w \bar\epsilon = 0$. Notice that $F_{12} = -2iF_{w\bar w}$. Tt will be convenient to write spinors in components, in particular the SUSY parameters are $\epsilon^+, \bar\epsilon^+$.
First we have a chiral multiplet $\Phi = (\phi, \bar\phi, \psi^-, \bar\psi^-)$ with variations
\bea
\label{SUSY var chiral}
\delta \phi &= -i \bar\epsilon^+ \psi^- \qquad\qquad & \delta \psi^- &= 2i \, \epsilon^+ D_{\bar w} \phi \\
\delta \bar\phi &= -i \epsilon^+ \bar\psi^- & \delta\bar\psi^- &= 2i \, \bar\epsilon^+D_{\bar w} \bar\phi \;.
\eea
Second we have a Fermi multiplet $\Lambda = (\psi^+, \bar\psi^+, G, \bar G)$ with variations
\bea
\label{SUSY var Fermi}
\delta \psi^+ &= \bar\epsilon^+ G + i \epsilon^+ E \qquad\qquad & \delta G &= 2\, \epsilon^+ D_{\bar w} \psi^+ - \epsilon^+ \psi_E^- \\
\delta \bar\psi^+ &= \epsilon^+ \bar G + i \bar\epsilon^+ \bar E & \delta \bar G &= 2\, \bar\epsilon^+ D_{\bar w} \bar\psi^+ - \bar\epsilon^+ \bar\psi_E^- \;.
\eea
Here $\cE(\Phi_i) = (E, \bar E, \psi_E^-, \bar\psi_E^-)$ is a chiral multiplet, holomorphic function of the fundamental chiral multiplets in the theory, and it is part of the definition of $\Lambda$. Notice that $E = E(\phi_i)$ and its fermionic partner is $\psi_E^- = \sum_i \psi_i^- \, \partial E/\partial \phi_i$. Third we have a vector multiplet \hbox{$V = (A_\mu, \lambda^+, \bar\lambda^+, D)$} with variations
\bea
\label{SUSY var vector}
\delta A_w &= \tfrac 12 \big( \epsilon^+ \bar\lambda^+ - \bar\epsilon^+ \lambda^+ \big) \qquad &
\delta \bar\lambda^+ &= \bar\epsilon^+ (-D-iF_{12}) \qquad &
\delta (-D-iF_{12}) &= 2\, \epsilon^+ D_{\bar w} \bar\lambda^+ \\
\delta A_{\bar w} &= 0 &
\delta \lambda^+ &= \epsilon^+ (-D+iF_{12}) &
\delta (-D+iF_{12}) &= 2\, \bar\epsilon^+ D_{\bar w} \lambda^+ \;.
\eea
Comparing with \eqref{SUSY var Fermi}, notice that the fields in the second and third column form a Fermi multiplet $\Upsilon = (\bar\lambda^+, \lambda^+, -D-iF_{12}, -D + iF_{12})$ with $\cE = 0$.

The supersymmetric action for chiral multiplets comes from the Lagrangian
\bea
\cL_\Phi &= D_\mu \bar\phi D^\mu \phi + i \bar\phi D \phi + 2\, \bar\psi^- D_w \psi^- - \bar\psi^- \lambda^+ \phi + \bar\phi \bar\lambda^+ \psi^- \\
&= -4 \bar\phi D_w D_{\bar w} \phi + \bar\phi (F_{12} + iD) \phi + 2\, \bar\psi^- D_w \psi^- - \bar\psi^- \lambda^+ \phi + \bar\phi \bar\lambda^+ \psi^- \;,
\eea
where the second equality is up to total derivatives.
For Fermi multiplets we have
\be
\cL_\Lambda = - 2\, \bar\psi^+ D_{\bar w} \psi^+ + \bar E E + \bar G G + \bar\psi^+ \psi_E^- - \bar\psi_E^- \psi^+
\ee
and for vector multiplets we have
\be
\cL_\Upsilon = \Tr \Big[ F_{12}^2 + D^2 - 2\, \bar\lambda^+ D_{\bar w} \lambda^+ \Big] \;.
\ee
Up to total derivatives, this equals the Lagrangian for the Fermi multiplet $\Upsilon$ with $\cE=0$.
Interactions are specified by holomorphic functions $J^a(\phi)$ of the chiral multiplets (and anti-holomorphic functions $\bar J^a(\bar\phi)$ of their partners), where $a$ parametrizes the Fermi multiplets in the theory:
\be
\cL_J = \sum\nolimits_a \big( G_a J^a + i \psi_a^+ \psi_J^{-a} \big) \;,\qquad\qquad \cL_{\bar J} = \sum\nolimits_a \big( \bar G_a \bar J^a + i \bar\psi_a^+ \bar\psi_J^{-a} \big) \;.
\ee
Their variation is a total derivative as long as
\be
\sum\nolimits_a E_a(\phi) J^a(\phi) = 0 \;.
\ee

All these actions are actually $\cQ$-exact. Let us define the anticommuting supercharge $\cQ$ by using commuting spinor parameters and choosing them $\epsilon^+ = \bar\epsilon^+ = 1$. The action of $\cQ$ is then immediately read off from (\ref{SUSY var chiral}), (\ref{SUSY var Fermi}) and (\ref{SUSY var vector}). We then find, up to total derivatives:
\bea
\cL_\Phi &= \cQ \big( 2i\bar\phi D_w \psi^- - i \bar\phi \lambda^+ \phi \big) \;,\qquad\qquad &
\cL_\Lambda &= \cQ \big( \bar\psi^+ G - i \bar E \psi^+ \big) \\
\cL_J &= \cQ \big( {\textstyle \sum_a} \psi_a^+ J^a \big) \;,\qquad\qquad &
\cL_\Upsilon &= - \cQ \, \Tr \big( \lambda^+ (D + iF_{12}) \big) \;.
\eea

In the reduction from $(2,2)$ to $(0,2)$ supersymmetry, the chiral multiplet $\Phi_{(2,2)}$ splits into a chiral multiplet $\Phi = (\phi, \bar\phi, P_-\psi, P_-\bar\psi)$ and a Fermi multiplet $\Lambda = (P_+\psi, P_+\bar\psi, F, \bar F)$. The vector multiplet $V_{(2,2)}$ splits into a vector multiplet $V$, with corresponding Fermi multiplet $\Upsilon = (P_+\bar\lambda, P_+\lambda, -D-F_{12}, -D+iF_{12})$, and an adjoint chiral multiplet $\Sigma = (\sigma, \bar\sigma, P_-\lambda, P_-\bar\lambda)$.
If $\Phi_{(2,2)}$ is charged under $V_{(2,2)}$, then its Fermi component $\Lambda$ has related chiral multiplet $\cE = \Sigma \Phi$ (where $\Sigma$ acts in the correct representation). It is easy to check that $\cL_\Upsilon + \cL_\Phi$ (where $\Phi$ is taken in the adjoint representation) equals $\cL_\text{YM}$, and $\cL_\Phi + \cL_\Lambda$ (where the Fermi multiplet has $\cE = \Sigma \Phi$) equals $\cL_\text{mat}$.
Superpotential interactions $W(\Phi_{(2,2)})$ become interactions $J^a(\phi) = \partial W/\partial \phi_a$.

Similarly, a $(2,2)$ twisted chiral multiplet $Y_{(2,2)}$ (which must be neutral) splits into a chiral and a Fermi multiplet. In particular the twisted chiral multiplet $\Sigma_{(2,2)}$ constructed out of $V_{(2,2)}$ splits into $\Upsilon$ (with $\cE=0$) and the chiral multiplet $\Sigma$. A twisted superpotential $\widetilde W(\Sigma_{(2,2)})$ becomes an interaction $J^\Upsilon(\sigma) = \partial \widetilde W / \partial \sigma$, and a complexified Fayet-Iliopoulos term is simply a constant $J^\Upsilon = \frac\theta{2\pi} + i\zeta$.

\bibliographystyle{ytphys}
\small\baselineskip=.93\baselineskip
\bibliography{ref}

\end{document}